\documentclass[12pt,letterpaper]{article}
 
\usepackage{amsmath}
\usepackage{amsfonts}
\usepackage{amssymb}

\usepackage{graphicx}

\def\be{\begin{equation}}
\def\ee{\end{equation}}
\def\ba{\begin{align}}
\def\ea{\end{align}}
\def\bear{\begin{eqnarray}}
\def\eear{\end{eqnarray}}
\def\nn{\nonumber}

\def\half{{{1\over 2}}}

\def\Heff{H_{\mathrm{eff}}}

\begin{document}


\begin{titlepage}
\vskip 1in
\begin{center}
{\Large
{Emergent geometry of membranes}}
\vskip 0.5in
{Mathias Hudoba de Badyn,
Joanna L. Karczmarek,  Philippe Sabella-Garnier and Ken Huai-Che Yeh}
\vskip 0.3in
{\it 
Department of Physics and Astronomy\\
University of British Columbia\\
Vancouver, Canada}
\end{center}

\vskip 0.5in
\begin{abstract}

In work \cite{Berenstein:2012ts}, a surface embedded in flat $\mathbb{R}^3$ is associated to 
any three hermitian matrices.  We study this emergent surface when the matrices
are large, by constructing coherent states corresponding to
points in the emergent geometry.  We find the original matrices determine not only
shape of the emergent surface, but also a unique Poisson structure. 
We prove that commutators of matrix operators correspond to Poisson brackets.
Through our construction, we can realize arbitrary noncommutative membranes:
for example, we examine a round sphere with a non-spherically symmetric Poisson structure.
We also give a natural construction for a noncommutative torus embedded in $\mathbb{R}^3$.
Finally, we make remarks about area and find matrix equations for minimal area surfaces.
\end{abstract}
\end{titlepage}

\tableofcontents


\section{Introduction}
\label{intro}

String theory contains many hints that spacetime might be a more
complicated object---possibly even an emergent one---than a manifold.
Most of our understanding about non-perturbative string theory comes
from the study of D-branes, extended objects that strings are
allowed to end on.  When $N$ identical D-branes are considered,
their coordinate positions are described by $N\times N$ hermitian
matrices.  If these matrix coordinates are simultaneously diagonalizable, 
their eigenvalues are easily interpreted as the positions of the D-branes.  
When they are not, as is the situation generically, the D-brane 
positions are not well defined, even in the classical 
$\hbar \rightarrow 0$ limit.  Thus, D-branes do not `view' spacetime
in the same way that ordinary point particles do.
The standard string theoretic interpretation of such `fuzzy' configurations
through the so-called dielectric effect \cite{Myers:1999ps},
where lower dimensional D-branes `blow up' to form higher dimensional
D-brane.  Lack of locality is related to the lower dimensional D-branes
being `smeared' over the worldvolume of a higher dimensional emergent
object.

In most previous work, explicit geometric interpretation of the matrix
coordinates as a higher dimensional object has been limited to simple and highly symmetric geometries, 
such as planes, tori and spheres.\footnote{One example of an attempt in 
a more general setup is \cite{Ellwood:2005yz}, where a matrix configuration
corresponding to a given surface was constructed using string boundary
states if zero energy states of a certain Hamiltonian arising from the 
boundary action can be found.}  In their paper, \cite{Berenstein:2012ts},
take this one step further: using the BFSS model
they found a geometric interpretation of three matrix coordinates
as a co-dimension one surface embedded in three dimension.
The argument was to consider a stack of D0-branes at an orbifold point, and 
then introduce an extra probe brane into the system.  By considering
a fermionic string stretching between the stack and the probe brane,
the emergent surface was defined as the locus of possible positions for the probe brane
where the stretched string has a massless mode (indicating
that the string has zero length).  This lead  to the following
effective Hamiltonian:
\be
\Heff(x_i) = \sum_{i=1,2,3} \sigma^i\otimes\left (X_i - x_i \right ) ~,
\label{BD}
\ee
where $X_i$ for $i=1,2,3$ are Hermitian, $N \times N$, matrices
corresponding to the positions of the stack of D0-branes in
a three dimensional flat transverse space, and $x_i$ are
the positions of the probe brane.  The fermionic mode is massless
when $\Heff$ has a zero eigenvalue.  Thus, the surface
corresponding to the three matrices $X_i$ is given by
the polynomial equation $\det(\Heff(x_i)) = 0$.  This defines a co-dimension one
surface in flat $\mathbb{R}^3$ space parametrized by $(x_1,x_2,x_3)$.

We use equation (\ref{BD}) as the starting point
for a concrete and explicit study of geometry of the emergent surface,
identifying zero eigenvectors of $\Heff$ with coherent
states underlying noncommutative geometry of the emergent surface
\cite{Berezin:1974du}.
We focus on configurations where a
smooth and well-defined surface arises from matrices with a large
size $N$.  Rather than assume it a priori, we {\it prove}
a correspondence principle between matrix commutators and
a unique Poisson bracket on the emergent surface arising from
the matrix configuration $(X_1,X_2,X_3)$.  This explicit correspondence
makes the usual procedure of going from matrix models to 
surfaces much less ad hoc, which might be of use when
quantizing membrane actions by replacing them with a matrix model.
We demostrate how easy it is to construct surfaces with desired
properties using our approach on several nontrivial examples, including
the torus.

Our approach is most similar to that espoused in 
\cite{Steinacker:2011ix} (see also \cite{Steinacker:2010rh} and references
therein), but with an explicit construction for the coherent
states associated with points on the surface.  
The results can also be thought of as a concrete realization of the abstract
idea in the classic work by Kontsevich, \cite{1997q.alg.....9040K}.  
Other work includes
\cite{Arnlind:2006ux,2007arXiv0711.2588A}, though 
our construction appears more general as it allows us to vary the local
noncommutativity independent of the shape of the surface.

For most of the paper, we focus on the following question:
under what conditions would a sequence of noncommutative
geometries, each arising from a matrix configuration $(X_1,X_2,X_3)$ and labeled by 
an increasing matrix size $N$, converge to a smooth limit?  
which quantities characterize the surface in this limit?

Since the polynomial equation $\det(\Heff(x_i)) = 0$ has degree $2N$, generically, the locus of its 
solutions does not need to be smooth in the large $N$ limit.
When some generic matrices $X_i$ are scaled so that the range of their
eigenvalue distributions remains finite at large $N$, the resulting
surface is generically quite complicated and does not have a large $N$ limit.
As a simple (but not generic) example, let
$X_i = \mathrm{diag}(\sigma_i + a_i^1, \ldots, \sigma_i + a_i^N)$,
where $\sigma_i$ are the Pauli matrices and $a_i^k$ are real numbers.  
The resulting surface
is a union of $N$ spheres of radius 1 each centered at $(a_1^k, a_2^k, a_3^k)$
for $k$ from 1 to $N$.  There is no sense in which the surface
achieves a well-defined large $N$ limit.  In the degenerate case where
all $a_i^k$ are zero, the surface is a single sphere of radius
one centered at the origin. However, it still does not correspond to a 
smooth geometry, rather, it
is a very fuzzy sphere with SU(N) symmetry.  
To obtain a smooth geometry, we can instead consider 
 $X_i = L_i/J$, with $L_i$ forming the irreducible
representation of SU(2) with spin $J$ (this is the standard construction
of the noncommutative sphere, see section \ref{subsection:sphere} for
details).  This sphere has radius $1$ independent of $J$.
As $N=2J+1 \rightarrow \infty$, the noncommutative sphere
reproduces the ordinary one.

When the the large $N$ limit exists and is smooth, the emergent surface will be characterized
by its geometry (the embedding into flat $\mathbb{R}^3$ space) and
by a Poisson structure defining (together with $N$) a noncommutative
geometry in the large $N$ limit.  In section \ref{setup}, we will make some
definitions and introduce our approach.  In section \ref{examples}, we will
analyze, analytically and numerically, a series of examples from which
a general picture will emerge.  In section \ref{largeN} we will prove
the correspondence principle and discuss smoothness conditions
which determine how large $N$ has to be for a given noncommutative
surface to be well described by the corresponding matrices.  In section \ref{sec:area},
we will discuss the issue of area and derive the matrix equation for 
minimal area surfaces. In section \ref{sec:torus}, we construct
a smooth torus embedded in $\mathbb{R}3$. Finally, in section \ref{future} we discuss 
topics for future work.

\section{Basic setup}
\label{setup}

Since our emergent surface is given by the locus of points where
the effective Hamiltonian $\Heff$ in equation (\ref{BD}) has
a zero eigenvalue, for each point $p$ on the surface
$\Heff$ has (properly normalized) zero eigenvector $|\Lambda_p\rangle$:
\be
\Heff |\Lambda_p \rangle = 0~.
\label{zero-eigenvector}
\ee
The above equation defines (in non-degenerate cases) a
two dimensional surface embedded in three dimensional space.
We will take the three dimensional space to be flat; the metric on the
emergent two dimensional surface will then just be the pullback from
the flat three dimensional metric.

It is instructive to rewrite the above equation in a slightly
different way:
\be
\sum_{i=1,2,3} \left (\sigma^i\otimes X_i  \right ) |\Lambda_p\rangle 
= \sum_{i=1,2,3} \left (\sigma^i\otimes x_i  \right ) |\Lambda_p\rangle ~.
\ee
This equation can be thought of as an analogue of an
eigenvalue equation: while the three matrices $X_i$ cannot be 
simultaneously diagonalized, the above equation says that 
if we double the dimensionality of the space under consideration,
there are special vectors $|\Lambda_p\rangle$ on which the action
of $X_i$ is described by only three parameters.  In analogy with the Berezin
approach to noncommutative geometry \cite{Berezin:1974du}, we would like to think of these
states as coherent states corresponding to points on the 
noncommutative surface.\footnote{A somewhat similar approach but
with a different effective Hamiltonian, and applicable only
in the infinite $N$ limit, was recently made in
\cite{Ishiki:2015saa}.}.
In the Berezin approach, every point $p$ is associated with a
coherent state $| \alpha_p \rangle$.  One then defines
a map from any  $\hat A$ to a function on the 
noncommutative surface via
$s(\hat A) = \langle \alpha_p | \hat A | \alpha_p \rangle$.
This function is usually called the symbol map.
From it one can find the corresponding star-product and the rest of the
usual machinery of noncommutative geometry.

The first difficulty we
see with $\Lambda_p$ being the coherent state is that our operators $X_i$ (and their functions) cannot
be seen as acting on $|\Lambda_p\rangle$ due to dimension mismatch.
We can simply `double' these operators by using $\boldsymbol{1}_2 \otimes X_i $
instead ($\boldsymbol{1}_k$ will denote the $k\times k$ identity matrix).
However, while it is true that
\be
\langle \Lambda_y |~\boldsymbol{1}_2 \otimes X_i ~| \Lambda_p \rangle
= x_i(\sigma) ~,
\ee
this approach is somewhat artificial.  We will see that
there is a more natural solution: for large $N$, when the emergent
noncommutative surface is smooth in the sense discussed in the Introduction,
the eigenvector $|\Lambda_p\rangle$ is approximately
a product, $|\Lambda_p\rangle = |a \rangle  \otimes  |\alpha_p\rangle$,  where
$|\alpha_p\rangle$ is $N$-dimensional and $|a\rangle$ is 2-dimensional.
In the next section, we will examine examples in which the 
zero eigenvectors of $\Heff$ do factorize in this manner when
$N$ is large.  A way to measure the extent of the factorization is to 
write any (2$N$)-dimensional vector as
\be
|\Lambda_p \rangle  =  \left [ \begin{array}{c} |\alpha_1\rangle \\ \hline 
|\alpha_2 \rangle \end{array}\right ]~,
\ee
with $ ||\alpha_1||^2 + ||\alpha_2||^2 = 1$, and to define
\be
A_p = \sqrt {|| \alpha_1 ||^2 || \alpha_2 ||^2 - |\langle \alpha_1 | \alpha_2 \rangle |^2}~,
\ee
which can be thought of as the area of the parallelogram defined by 
the two vectors $|\alpha_1\rangle$ and $|\alpha_2\rangle$.  
We will be arguing that, in the large $N$ limit, $A_p$ is of order 
$N^{-1/2}$, implying that $|\alpha_1\rangle$ and $|\alpha_2\rangle$ are
indeed approximately parallel and we can write 
\be
|\Lambda_p \rangle  =  \left [ \begin{array}{c} a |\alpha_p\rangle \\ \hline 
b |\alpha_p \rangle \end{array}\right ] + \mathcal{O}(1/\sqrt{N})~.
\label{lambda-factorized}
\ee
(By $\mathcal{O}(1/N^{-1/2})$ we mean that the norm of the correction vector 
decreases with increasing $N$ like $1/N^{-1/2}$.)
It will then be the $N$-dimensional vector $|\alpha_p\rangle$ that will
play the role of a coherent state corresponding to point $p$.

The complex coefficients $(a,b)$ of the 2-vector $|a\rangle$ determine
the direction of the normal vector $n$ at point $p$ given by $(x_1,x_2,x_3)$. 
To see this, consider moving $p$ slightly to $(x_1+dx_1,x_2+dx_2,x_2+dx_3)$, where
$(dx_1,dx_2,dx_3)$ is an infinitesimal tangent to the surface.   First order
perturbation theory implies that to maintain the condition
that $\Heff$ has a zero eigenvalue, we must have 
$\langle \Lambda_p |d\Heff |\Lambda_p \rangle = \langle \Lambda_p |  \sigma^i\otimes (-dx_i) |\Lambda_p \rangle = 0$.
Thus $dx_i~ \langle \Lambda_p |\sigma^i\otimes{\mathbf 1}_N  |\Lambda_p \rangle = 0$,
implying that 
\be
n_i := \langle \Lambda_p |~ \sigma^i\otimes {\mathbf 1}_N ~|\Lambda_p \rangle
\label{normal}
\ee
is a vector  normal to the surface at a point $p$.
This is an exact statement and does not rely on our factorization assumption.
Incidentally, we have the formula $|n|^2 =  1 - 4A_p^2$, so the normal vector is
close to being a unit normal when the factorization condition holds.
When we use equation (\ref{lambda-factorized}), we obtain that
the normal vector is 
$(\bar a b + a \bar b, i(a\bar  b -  \bar a b), \bar a a - \bar b b)$.
Thus, the coefficients $(a,b)$ fix the direction of the normal vector. 
Conversely, the normal vector fixes the coefficients $(a,b)$ up to an
overall irrelevant phase.

Next, we will try to define local noncommutativity on the surface.
The local noncommutativity can be thought of in two different ways:
the size of `fuzziness' (or uncertainty) of the operators $X_i$
in the state $|\Lambda_p\rangle$, or the size of the commutators
of the $X_i$s when acting on $|\Lambda_p\rangle$.  In a coherent state,
these two notions should be equal, and they turn out to be equal here,
strengthening our case that $|\Lambda_p\rangle$ can be thought of as a coherent
state.  Using $\sigma^i \sigma^j = i \epsilon^{ijk} \sigma^k
= - \sigma^j \sigma^i$ for $i\neq j$ and $\sigma_i^2 = 1$, we have
a nice little identity
\be
(\Heff)^2 = \boldsymbol{1}_2\otimes\sum_i (X_i- x_i)^2 
~ + ~\half i \epsilon_{ijk}\sigma_i\otimes[X_j,X_k]~.
\ee
Then, since $\langle \Lambda_p | (\Heff)^2 |\Lambda_p \rangle = 0$, 
we have
\be
\langle \Lambda_p|\boldsymbol{1}_2 \otimes   \sum_i (X_i- x_i)^2 
| \Lambda_p \rangle = 
- \half i \epsilon_{ijk}~ \langle \Lambda_p |\sigma_i\otimes[X_j,X_k]| \Lambda_p \rangle~. 
\ee
When the vector $|\Lambda\rangle$ is indeed a product, we can use equation
(\ref{lambda-factorized}) to make the following definition: the local noncommutativity
on the noncommutative surface is
\be
\theta =  \langle \alpha_p | \sum_i (X_i- x_i)^2 | \alpha_p \rangle = 
 \half \epsilon_{ijk}~ \theta_{ij} ~ n^k~,
\label{noncommutativity}
\ee
where
\be
\theta_{ij} := \langle \alpha_p | -i [X_i,X_j] | \alpha_p \rangle~.
\ee

The LHS of expression (\ref{noncommutativity}) is a sum of squares of uncertainties
in the operators $X_i$, while the RHS depends on the commutators.
The particular combination of commutators is of interest:
with our factorization assumption, the commutator term picks up only
the contributions that are transverse to the normal, for example, if the normal
vector $n$ is pointing in the $x_3$ direction, only $[X_1,X_2]$ contribute to $\theta$.
In fact, it will turn out that, in the large $N$ limit, 
$\epsilon_{ijk}\theta_{ij}$ is nearly parallel to $n_k$.
Thus, we can also write $\theta$ as
\be
\theta = \langle \alpha_p| ~
 \sqrt{\sum_{i\neq j}-[X_i,X_j]^2}~|\alpha_p \rangle~.
\label{noncommutativity-2}
\ee

As for the first expression in equation (\ref{noncommutativity}), it will turn
out that if we take the normal vector to point along the $x_3$ direction, we have
$\langle \alpha | (X_1- x_1)^2 | \alpha \rangle \approx
\langle \alpha | (X_2- x_2)^2 | \alpha \rangle \gg
\langle \alpha | (X_3- x_3)^2 | \alpha \rangle$, so the coherent state is
`flattened' to lie predominantly in the 1-2-plane and balanced (`round').

To flesh out these ideas, we will examine a series of increasingly complex examples.
In the process, we will construct the approximate eigenvector $|\alpha_p\rangle$
and study corrections to the large $N$ limit described above.

\section{Coherent state and its properties}
\label{examples}

We will make the following choice for the Pauli matrices $\sigma_i$
\be
\sigma^1 = \begin{bmatrix}
0 & 1 \\ 1 &0
\end{bmatrix}~,~~~
\sigma^2 = \begin{bmatrix}
0 & -i \\ i &0
\end{bmatrix}~,~~~
\sigma^3 = \begin{bmatrix}
1 & 0 \\ 0 &-1
\end{bmatrix}~.
\ee
In this convention, we can write $\Heff$ in a natural way in terms of $N \times N$ blocks
\be
\Heff = \left [ \begin{array}{c|c}
X_3 - x_3 & (X_1-iX_2) - (x_1 - ix_2) \\ \hline 
(X_1+iX_2) - (x_1 + ix_2)  & -(X_3 - x_3)
\end{array} \right ]~,
\ee

We will now examine a series of examples of increasing
complexity, always focusing on a point where the normal
vector to the surface is pointing straight up (in the $x_3$ direction).

Our final conclusion will be that at such a point, the zero-eigenvector
of $\Heff$ has the form given in equation (\ref{lambda-factorized}):
\be
|\Lambda \rangle  =  \left [ \begin{array}{c} |\alpha\rangle \\ \hline 0
\end{array}\right ]~+~{\cal O}\left ( N^{-1/2}\right ).
\label{largeN-eigenvector}
\ee
$|\alpha \rangle$  with $\langle \alpha | \alpha \rangle = 1$ will be
the coherent state associated with this particular point on the surface,
 $-i\langle \alpha |[X_1,X_2]| \alpha \rangle$ will correspond
to the local value of noncommutativity at this point.
This result is easily generalizable to any orientation of  the surface
using an SU(2) rotation of the Pauli matrices.

\subsection{Example: noncommutative plane}
\label{subsection:ncplane}

Consider the example of a noncommutative plane: let $X_3 = 0$,
and let $[X_1, X_2] = i\theta$.  Out of necessity, $X_1$ and $X_2$ are
infinite dimensional operators.  This will not be the case
when we are considering compact noncommutative surfaces.
We have
\be
\Heff = \left [ \begin{array}{c|c}
- x_3 & A^\dagger - \bar \alpha  \\ \hline A - \alpha  & x_3
\end{array} \right ]~,
\ee
where $A = X_1 + i X_2$,
$A$ and $A^\dagger$ are the lowering and raising operators
of a harmonic oscillator with $[A,  A^\dagger] = 2\theta$, 
and $\alpha = x_1 +i x_2$.
The lowering operator $A$ has
eigenstates $|\alpha \rangle$, called the coherent states,
corresponding to  every complex number $\alpha$: 
$A |\alpha \rangle =  \alpha |\alpha \rangle$.  We thus
have a zero eigenvector for $\Heff$ with $x_3=0$: 
\be
|\Lambda(\alpha) \rangle  =  \left [ \begin{array}{c} |\alpha\rangle \\ \hline 0
\end{array}\right ]~.
\ee

The noncommutative plane is flat and has constant noncommutativity.
The normal vector is $
\langle \Lambda | \sigma^i \otimes {\mathbf 1} |\Lambda \rangle = (0,0,1)$ and 
we have $-i\langle \alpha [X_1, X_2]  \alpha \rangle = \theta$.

The importance of this example is that, locally and in the large $N$ limit, 
any noncommutative surface should look like the 
noncommutative plane.  This is the observation that
will allow us to write our definition of a large $N$ (smooth) limit.




\subsection{Example: noncommutative sphere}
\label{subsection:sphere}

Here we have $X_i = L_i/J$ where $L_i$ form the $N$-dimensional
irrep of SU(2): $[L_i, L_j] =  i \epsilon_{ijk} L_k$ and
where $J=(N-1)/2$ is the spin.  It is useful to consider
the usual raising and lowering operators, $L_\pm = L_1 \pm i L_2$.
Without loss of generality,
consider that point on the noncommutative surface which
lies on the $x_3$ axis.  With $x_1 = x_2 = 0$, $\Heff$ is
\be
\Heff = \left [ \begin{array}{c|c}
L_3/J - x_3 & L_-/J  \\ \hline L_+/J  &-(L_3/J-x_3)
\end{array} \right ]~. 
\ee
We will use as a basis the eigenvectors of the $L_3$ 
angular momentum, $|m\rangle$:
\be
L_3 |m\rangle = m|m\rangle~,~~m=-J\ldots J~,~~ \langle  m | m \rangle =1~,
~~J=\frac{N-1}{2}~.
\ee
It is easy to see that
\be
|\Lambda \rangle = \left [ \begin{array}{c} |J\rangle \\ \hline 0
\end{array}\right ]
\label{coherent-state-sphere}
\ee
is a zero eigenvector of $\Heff$ if $x_3 = 1$.  Thus, the
noncommutative sphere has radius 1.\footnote{
This is a different definition of the radius of the noncommutative
sphere than the usual one, which is based on the quadratic Casimir
of the SU(2) irrep, and which gives the radius to be $\sqrt{N^2-1}/J
= \sqrt{\frac{N+1}{N-1}}$.
}

\subsection{Looking ahead: polynomial maps from the sphere}
\label{subsection:variety}

A large class of surfaces that can be studied using our
tools are surfaces that are generated from polynomials 
of the normalized SU(2) generators considered above:
\be
X_i = \mathrm{polynomial}(L_1/J, L_2/J, L_3/J)~,
\ee
where the  polynomials in three variables have
degrees and coefficients that are independent of $N$.  In this
case, we expect that at large $N$ the noncommutative
surface will approach an algebraic variety given by
the image of the unit sphere under the polynomial maps used
to construct $X_i$.

Concretely, consider a surface $\cal{S}$ 
in $\mathbb{R}^3$ constructed as follows: let $p_1$, $p_2$ and $p_3$
be three polynomials discussed, in three variables $w_1$, $w_2$ and $w_3$.  
Then, consider the image in $\mathbb{R}^3$ under these three polynomial maps
of the  surface $\sum_i (w_i)^2 = 1$, ie
\be
{\cal S} = \left \{ (x_1, x_2, x_3) ~|~x_i = p_i(w_1, w_2, w_3)~
\mathrm{and}~\sum_i (w_i)^2 = 1 \right \}~.
\ee  We will restrict our considerations to surfaces
which are non-self-intersecting, meaning that the polynomial
map is one-to-one.  The corresponding noncommutative surface 
is specified by three $N \times N$ matrices $X_i$ which can be 
written as corresponding polynomial expressions in $L_i$:
\be
X_i = \mathrm{sym}\left (  p_i( L_1/J, L_2/J,  L_3/J ) \right )~,
\ee
where, to avoid ambiguity, 
the `sym' map completely symmetrizes any products of
the three non-commuting matrices $L_i$.  This symmetrization
will turn out to play little role in what follows: re-ordering
the terms of order $k$ leads to small---suppressed by a
power of $J$---corrections in the
coefficients of the polynomials of order less than $k$.

Now, consider an arbitrary point $p = (y_1, y_2, y_3)$ 
on the surface $\cal{S}$. 
Acting with SO(3) on the space $(x_1, x_2, x_3)$, arrange for the 
normal vector to $\cal{S}$ at the point $p$ to point along the positive
$x_3$-direction, and acting with SO(3) on the space $(w_1, w_2, w_3)$, 
arrange for the pre-image of the point $p$ to be the north pole.
It is then necessary that the polynomial maps take a form
\begin{align}
x_1 &~=~ y_1 + c_1 w_1 + c_2 w_2 &+~ a (w_3-1)~+~p_1^{(2)}(w_1, w_2, w_3-1)~,&  \nn \\
x_2 &~=~ y_2  + c_3 w_1 + c_4 w_2 &+~ b (w_3-1)~+~p_2^{(2)}(w_1, w_2, w_3-1)~,& \label{polynomial}\\
x_3 &~=~ y_3                      &+~  c (w_3-1) ~+~p_3^{(2)}(w_1, w_2, w_3-1)~,& \nn
\end{align}
where $c_i$, $a$, $b$ and $c$ are real numbers and where
$p_i^{(2)}(\cdot)$ are polynomials of degree at least 2.
To avoid a coordinate singularity, we should have $c_1c_4 - c_2c_3 \neq 0$.
Then, using a rotation of $w_1$ and $w_2$ (in other words, rotating
the unit sphere around the north pole), we can set $c_3$ zero
and $c_4>0$.  Finally we can take $c_1 >0$ by adjusting the sign of
$w_1$ if necessary.  

The four coefficients $c_1, \ldots, c_4$ determine the metric on the 
surface in terms of the metric on the sphere.  If the metric on the 
sphere is $g_{S^2}$, then the induced metric on the surface is
\be
g_{ab}:= \left ({\mathbf C}^T g_{S^2}{\mathbf C}\right )_{ab}~,~
\mathrm{where}~ {\mathbf C} = \left [ \begin{array}{cc}c_1&c_2\\c_3&c_4\end{array}\right ]~.
\label{c-matrix-metric}
\ee
This implies that $\sqrt{\det g}/\sqrt{\det g_{S^2}} = \det {\mathbf C}$,
which is a useful fact to keep in mind.

Without loss of generality, we are interested in the eigenvector
of $\Heff$ at a point such that the normal to the surface is 
pointing along the 3-direction.  We now want to show that the 
corresponding zero-eigenvector of $\Heff$ has the form shown in equation
(\ref{largeN-eigenvector}).

Before we plunge into analyzing this rather general setup, we will
narrow the example down to a simpler one which nonetheless contains
most of the salient features of our general approach.

\subsection{Example: noncommutative ellipsoid}
\label{subsection:ellipsoid}

Here, we will consider a stretched noncommutative sphere.  The most 
generic closed quadratic surface in three dimensions is 
an ellipsoid, with three orthogonal major axes positioned
at some arbitrary position in the three dimensional space under consideration.
In other words, we will allow $X_i$ to be 
arbitrary linear combinations of $L_1/J$, $L_2/J$ and $L_3/J$.
Under the general framework described above, this amounts
to setting the higher degree polynomials $p_i^{(2)}$ to zero:
\be
X_i = \mathbf{A}_{ij} L_j/J,~~\mathrm{where~~}
\mathbf{A} = \left [ \begin{array}{ccc}
c_1 & c_2 & a \\
0 & c_4 & b \\
0 & 0 & c
\end{array} \right ]~.
\label{ellipsoid-def}
\ee

The classical, or infinite $N$, surface is given by 
$x_i = \mathbf{A}_{ij} w_j$ with $\sum_i (w_i)^2 = 1$.
It is easy to check that at a point $x = (a,b,c)$, this surface
has a normal vector which is pointing along the positive $x_3$-direction.
We will therefore consider finding the exact location of the 
surface at a point with $x = (a, b, x_3)$ where we expect $z_3$ to be 
close to c.  We have
\be
\Heff(x_3) = \left [ \begin{array}{c|c}
c\frac{L_3}{J}-x_3 &
A^\dagger ~+~(a-ib)(L_3/J - 1)
\\ \hline 
A ~+~(a+ib)(L_3/J - 1)  &-\left(c\frac{L_3}{J}-x_3
\right)
\end{array} \right ]~,
\ee
where 
\be
A = \frac{(c_1+c_4)-ic_2}{2J}L_+ + \frac{(c_1-c_4)+ic_2}{2J}L_-~.
\label{A}
\ee

What we need to do is find a good approximation to the zero eigenvector
of $\Heff(x_3)$, together with an estimate for the (hopefully small) difference
$x_3-c$.  We conjecture that such a vector is in some way similar to that in
equation (\ref{coherent-state-sphere}): the `top part' is large compared with
the `bottom part' and is dominated by components with the largest
eigenvalues of $J_3$.  To achieve this,  write $\Heff$ as a sum of two parts:
\be
\Heff(x_3) = \left [ \begin{array}{c|c}
 0 & A^\dagger
\\ \hline 
A & 0
\end{array} \right ]~
+~
\left [ \begin{array}{c|c}
c\frac{L_3}{J}  - x_3 & (a-ib)\left(\frac{L_3}{J} -1\right)
\\ \hline 
(a+ib)\left(\frac{L_3}{J} -1\right)  &-\left(c\frac{L_3}{J} -x_3\right)
\end{array} \right ]~.
\label{heff-two-parts-ellipsoid}
\ee
If we focus on vectors whose $N$-dimensional sub-vectors are
dominated by components with large $L_3$ eigenvalues, then the
first part can be thought of as being of order $N^{-1/2}$ 
while the second part is of order $N^{-1}$.  Our attempt
to find an approximate eigenvector of $\Heff(x_3)$ will treat
the second part as a small perturbation on the first part,
suppressed by $N^{-1/2}$.

Consider now a vector---which we will show to be 
either a zero eigenvector of $A$ or very close to such, and which
will thus be 
an approximate zero-eigenvector of $\Heff(x_3)$---given by
\be
\left [ \begin{array}{c} |\alpha\rangle \\ \hline 0
\end{array}\right ]~,
\label{eigenvector-leading-order}
\ee
where\footnote{Some standard notation we will 
use: the `floor' function, $\lfloor x \rfloor$ =  the largest integer
not exceeding $x$; the double factorial, $(2n)!! = (2n)(2n-2)\ldots(4)(2)$ and
$(2n-1)!! = (2n-1)(2n-3)\ldots(3)(1)$ for $n$ a natural number.}
\be 
|\alpha\rangle =  \frac{1}{\sqrt K}\sum_{m=0}^{\lfloor J \rfloor} 
~ \left ( \xi ^m~
\sqrt{\prod_{k=1}^m \frac
{(2k-1)(2J-2k+2)}
{(2k)(2J-2k+1)}}
\right ) ~|J-2m\rangle~,
\label{alpha}
\ee
with $\xi$ is given by
\be
\xi = - \frac{c_1 - c_4 + ic_2} {c_1 + c_4 - ic_2} ~.
\label{xi}
\ee
The normalization constant, for which $\langle \alpha | \alpha \rangle=1$, 
can be computed in the large $J$ limit as
\bear
K  &=&
\sum_{m=0}^{\lfloor J \rfloor} 
~ \left ( \left |\xi \right |^{2m}~
\prod_{k=1}^m \frac
{(2k-1)(2J-2k+2)}
{(2k)(2J-2k+1)}
\right )   \\ 
&\approx &  1 + \sum_{m=1}^{\infty} 
~ \left ( \left | \xi \right |^{2m}~
\frac {(2m-1)!!}{(2m)!!} 
\right ) ~~= ~~ 1+
2  \sum_{m=1}^{\infty} |\xi/2|^{2m} \frac {(2m-1)!}{m!(m-1)!} 
\\ &=& 1 + \frac{|\xi|^2}{1-|\xi|^2 + \sqrt{1-|\xi|^2}}
~~=~\frac{1}{\sqrt{1-|\xi|^2}}~,
 \eear
where it is important that $|\xi|<1$, which can be seen from the 
explicit form in equation (\ref{xi}).  For completeness, let us state
that 
\be
1-|\xi|^2 = \frac{4\det {\mathbf C}}{\|{\mathbf C}\|^2 + 2 \det {\mathbf{C}}} ~,
\ee
or
\be
\frac{1-|\xi|^2}{1+|\xi|^2} = \frac{2\det {\mathbf C}}{\|{\mathbf C}\|^2}~.
\ee
Writing $|\xi|$ in terms of rotational invariants of the matrix $\mathbf C$ gives a clear
geometric interpretation this is quantity: it is a measure of how much the map in equation
(\ref{ellipsoid-def}) distorts the aspect ratio at the point we are interested in.

With a short calculation\footnote{
Recall that
\be
 L_- | k \rangle = \sqrt{(J-k+1)(J+k)} ~ | k -1\rangle~,
~~~
 L_+ | k \rangle = \sqrt{(J-k)(J+k+1)} ~| k +1\rangle~.
\ee
}
we see that $A | \alpha \rangle = 0$ for integer spin $J$, and that
for half-integer spin $J$, we have
\bear
A | \alpha \rangle &=& -\frac{c_1 - c_4 + ic_2}{2J \sqrt K}
\left ( \xi^{J+1/2} 
\sqrt{ \prod_{k=1}^{J-1/2} \frac
{(2k-1)(2J-2k+2)}
{(2k)(2J-2k+1)}}
\right )
\sqrt{2J} ~|-J\rangle
\\ &=& K^{-1/2} (c_1 - c_4 + ic_2)~ \xi^{J+1/2}~  \frac {(2J-2)!!} {(2J-1)!!}
 ~|-J\rangle~. \eear 
This is very small: the norm-squared of 
$A | \alpha \rangle$ is  bounded above by
\be
b(J) := 
\left ((c_1-c_4)^2+(c_2)^2 \right ) \left | \xi\right |^{2J+1}~.
\ee
Since $|\xi|<1$, the above quantity goes to zero like $\exp(-(2\ln|\xi|) J)$ 
for large $J$.
Further, 
\be
 \left(\frac{L^3}{J} -1 \right )|\alpha\rangle
=  -\frac{1}{\sqrt K} \sum_{m=0}^{\lfloor J \rfloor} ~\frac {2m}{J}
~ \left ( \xi^m~
\sqrt{\prod_{k=1}^m \frac
{(2k-1)(2J-2k+2)}
{(2k)(2J-2k+1)}}
\right ) ~|J-2m\rangle
\ee
and the norm-squared of this vector is equal to
\be
\frac{1}{K}\sum_{m=0}^{\lfloor J \rfloor} ~\left (\frac {2m}{J}\right )^2
~ \left ( \left | \xi \right |^{2m}~
\prod_{k=1}^m \frac
{(2k-1)(2J-2k+2)}
{(2k)(2J-2k+1)}
\right )~,
\ee
which is bounded above by\footnote{
We need to provide a bound on
\be
\prod_{k=1}^m \frac {(2k-1)(2J-2k+2)} {(2k)(2J-2k+1)}
\ee
Consider, for $m$ a positive integer less or equal than $\lfloor J \rfloor$,
\be
F(m) := \prod_{k=1}^m \frac {(2k-1)(2J-2k+2)} {(2k)(2J-2k+1)}
= \frac {(2m-1)!!(2J-2m-1)!!(2J)!!} {(2m)!!(2J-2m)!!(2J-1)!!} ~.
\ee
$F(1) = \frac{J}{2J-1}<1$ and $F(\lfloor J \rfloor)$ can also be 
easily shown to be less than $1$ (we need to consider two cases, with
$J$ integer or half-integer).  Finally, we notice that
$F(m+1) < F(m)$ for $m$ smaller than roughly $J/2$ and 
$F(m+1) > F(m)$ for $m$ larger than than.  This implies that $F(m)$ 
has a minimum near $J/2$ and that for $1<m<\lfloor J \rfloor$ it is
less than the larger of $F(1)$ and $F(\lfloor J \rfloor)$ which are both less
than 1.  Therefore,
\be
\prod_{k=1}^m \frac {(2k-1)(2J-2k+2)} {(2k)(2J-2k+1)} < 1~.
\ee
}
\be
 \sum_{m=0}^{\lfloor J \rfloor} ~\left (\frac {2m}{J}\right )^2
~ \left ( \left | \xi \right |^{2m} \right ) 
<
J^{-2} \sum_{m=0}^{\infty} ~\left ({2m}\right )^2
~ \left ( \left | \xi \right |^{2m} \right ) 
:=  u(J)~.
\ee
Thus, the bound has the form 
$u(J) = (\mathrm{function~of~} \xi) \cdot J^{-2}$.

When $\Heff(x_3=c)$  acts on the normalized
vector $\left [ \begin{array}{c} |\alpha\rangle \\ \hline 0 \end{array}\right ]$, 
the resulting vector's norm is,
in the large $J$ limit, bounded by  $\sqrt{(a^2+b^2+c^2) u(J)+b(J)}$,
which is itself bounded by a constant times $J^{-1}$.  
To summarize,
\be
\left \| \Heff(c) \left [ \begin{array}{c} |\alpha\rangle \\ \hline 0
\end{array}\right ] \right \| ~<~ \frac{C(c_i)}{J}~,
\label{bound}
\ee
where $C(c_i)$ does not depend on $J$ and therefore on $N$.

It follows that
$\left [ \begin{array}{c} |\alpha\rangle \\ \hline 0 \end{array}\right ]$ 
is an approximate eigenvector of
$\Heff(c)$ and we can place a bound on the corresponding eigenvalue:
there exists a vector $\tilde \Lambda$ such that
\be
\Heff(c) ~\tilde \Lambda = \epsilon \tilde \Lambda~,~~\mathrm{with}~~
|\epsilon| <  \frac{C(c_i)}{J}~.
\label{epsilon}
\ee
One can ask the following question: is $\tilde \Lambda$ close to 
$\left [ \begin{array}{c} |\alpha\rangle \\ \hline 0 \end{array}\right ]$?
To answer this question, we examine the argument that guarantees
the existence of $\tilde \Lambda$ as above: consider the 
length squared of 
$\Heff \left [ \begin{array}{c} |\alpha\rangle \\ \hline 0 \end{array}\right ]$
as expanded in eigenvectors of $\Heff$:
\be
\Heff \Lambda_i = \lambda_i \Lambda_i~,~~~~
\Heff(c) \left [ \begin{array}{c} |\alpha\rangle \\ \hline 0
\end{array}\right ] = \sum_{i=1}^{2N} ~ c_i \Lambda_i
~,~~~~
\left \| \Heff(c) \left [ \begin{array}{c} |\alpha\rangle \\ \hline 0
\end{array}\right ] \right \|^2 = 
\sum_{i=1}^{2N} ~ |c_i|^2 |\lambda_i|^2~.
\ee
With the bound in equation (\ref{bound}), it is clear that at least
one of the eigenvalues $\lambda_i$ must be less than $C(c_i)/J$.
Further, if none of the other eigenvalues are small enough, then
the eigenvector corresponding to the unique small eigenvalue 
(which we denoted with $\tilde \Lambda$) is very close to 
$\left [ \begin{array}{c} |\alpha\rangle \\ \hline 0 \end{array}\right ]$ 
itself.  For example, if the next smallest eigenvalue $\lambda_j$ of 
$\Heff$ is of order $N^{-1/2}$ (as numerical studies suggest), 
then the corresponding coefficient $c_j$ must be of order
$N^{-1/2}$ as well.  Therefore, the difference between $\tilde \Lambda$
and $\left [ \begin{array}{c} |\alpha\rangle \\ \hline 0 \end{array}\right ]$ 
has length of order $N^{-1/2}$.

Further, we would like to conclude that there exists a third
vector $\Lambda$, such that
\be
\Heff(c-\zeta)~  \Lambda = 0~,~~\mathrm{with}~~
|\zeta| ~\mathrm{of~order}~1/J~,
\ee
with $\Lambda$  close to $\tilde  \Lambda$ and therefore 
$\left [ \begin{array}{c} |\alpha\rangle \\ \hline 0 \end{array}\right ]$.
It is possible to argue for this in first order perturbation theory:
as we deform $x_3$ from $c$ to $c-\zeta$, the eigenvalue of
interest changes from $\epsilon$ (in equation (\ref{epsilon})) to 0,
while the eigenvector changes from $\tilde \Lambda$ to
$\Lambda$.  Since
$\epsilon$ is of order $N^{-1}$, $\zeta$ should also be of
order $N^{-1}$. Making this analysis rigorous is difficult because, 
effectively, we are
trying to do perturbation theory in $1/N$ while taking a large $N$ limit.
Since any sums we take would be over $N$ components, these sums can
easily overwhelm any $1/N$ suppression factors.  
For example, to show that  $\Lambda$ is close to $\tilde \Lambda$, it is
again necessary to bound the remaining spectrum of $\Heff(c)$ away from zero. 
This is the same bound as was necessary above: 
the remaining eigenvalues must be bounded away from zero
by at least const$/\sqrt N$, which seems to be the case when
examined numerically.

Instead of attempting a rigorous proof, we
will obtain some analytic estimates based on
the assumption that the $1/N$ expansion is valid and then 
confirm these estimates with numerical analysis.

Our idea will be to obtain an analytic result for the leading
order contribution to $x_3-c$ (which will turn out to be of order $1/N$
as predicted above) and confirm its correctness by comparing with with numerical results.
We will also confirm that our approximate eigenvector 
$\left [ \begin{array}{c} |\alpha\rangle \\ \hline 0 \end{array}\right ]$
is a good approximation to the exact zero eigenvector of $\Heff(x_3)$.
Crucial to this approach are two facts: that the eigenvector $|\alpha \rangle$ 
has components which fall off exponentially with $m$, so that only those
components with spin close to the maximum spin $J$ are appreciable, and that
the second term in equation (\ref{heff-two-parts-ellipsoid}) is small
(of order $1/N$) when acting on these components.  Further
analysis will then reveal that when the first order correction to
the approximate eigenstate is included: 
$\left [ \begin{array}{c} |\alpha\rangle \\ \hline |\beta\rangle \end{array}\right ]$,
the vector $|\beta\rangle$ also has components which fall off exponentially with $m$.
We will interpret this as a `quasi-locality' feature of the noncommutative
surface.

\begin{figure}
\includegraphics[width=6.5in]{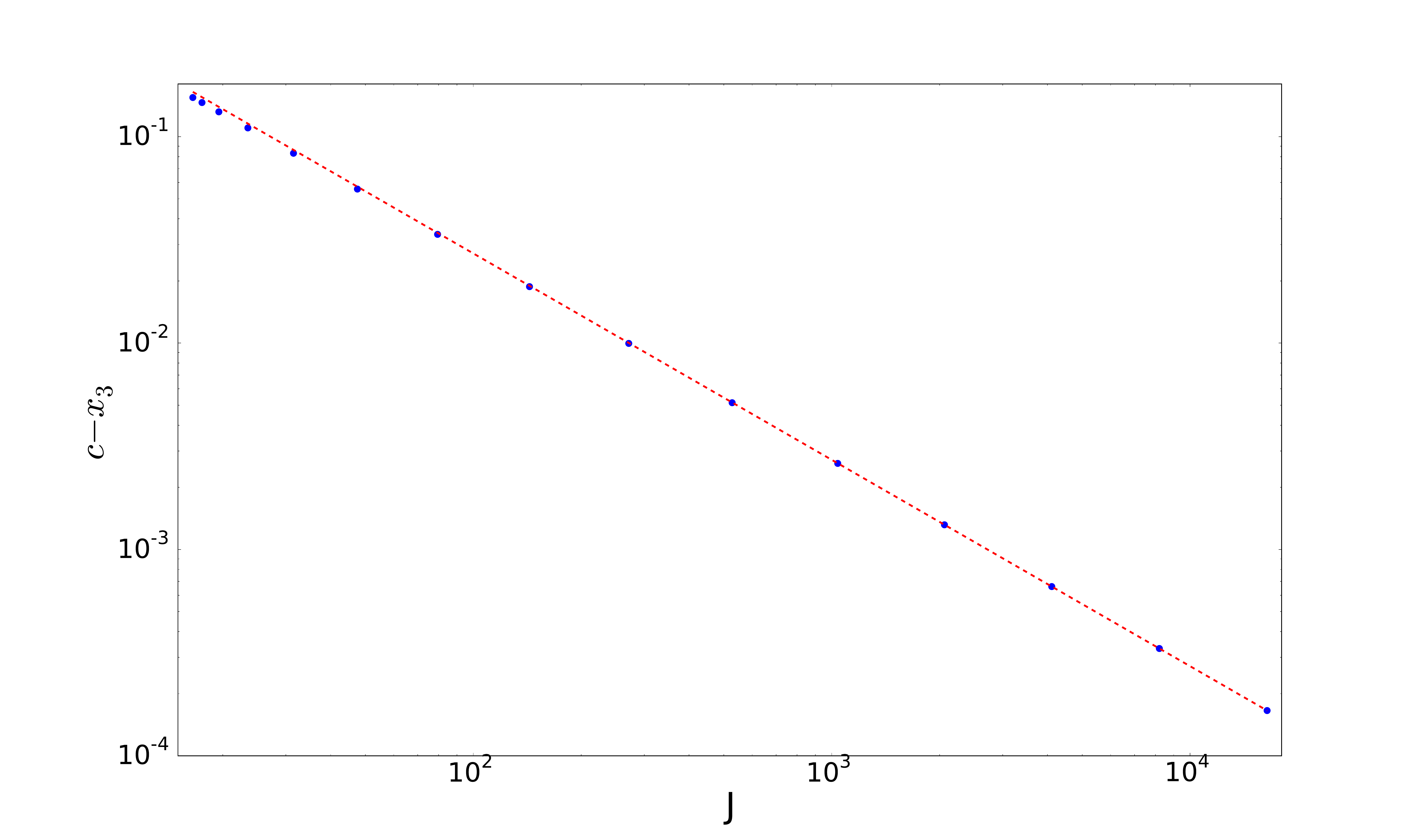}
\caption{Difference between $x_3$ at finite $N$ (obtained
numerically) and $c$ (its large $N$ asymptotics), as a function
of $N$.  The line represents equation (\ref{eq-correction-ellipsoid}),
which has no free parameters and appears to be an excellent match to the 
numerical data.  In this figure,
$(a,b,c)=(1.5,0.5,3)$, $c_1=2$, $c_2=5$ and $c_4=4$. For these
values, equation (\ref{eq-correction-ellipsoid}) implies that
$c-x_3 = 2.71875/J$.}
\label{f1}
\end{figure}

Now, return to our way of writing $\Heff$ as a sum of two
parts in equation (\ref{heff-two-parts-ellipsoid}).
Our special vector 
$\left [ \begin{array}{c} |\alpha\rangle \\ \hline 0 \end{array}\right ]$
is an approximate zero eigenvector of the first of these two operators
(and an exact zero eigenvector for odd N).  Thinking of the second
term in equation  (\ref{heff-two-parts-ellipsoid}) as a small perturbation
in first order perturbation theory, we obtain, to first order, that the 
change in the eigenvalue is equal to
\bear
&&
\left [ \begin{array}{c|c} \langle\alpha| & 0 \end{array}\right ]
\left [ \begin{array}{c|c}
c\frac{L_3}{J}  - x_3 & (a-ib)\left(\frac{L_3}{J} -1\right)
\\ \hline 
(a+ib)\left(\frac{L_3}{J} -1\right)  &-\left(c\frac{L_3}{J} -x_3\right)
\end{array} \right ]
\left [ \begin{array}{c} |\alpha\rangle \\ \hline 0 \end{array}\right ] \\ \nn &=& 
 \langle\alpha|  \frac{L_3}{J} c - x_3 |\alpha\rangle 
\\\nn &=& \langle\alpha|  \frac{L_3}{J} - 1 |\alpha\rangle c  ~+~
\langle \alpha | \alpha \rangle (c-x_3) \\ \nn &=&
- \frac{c}{K} \sum_{m=0}^{\lfloor J \rfloor} ~\frac {2m}{J}
~ \left ( \left |\xi \right |^{2m}~
\prod_{k=1}^m \frac
{(2k-1)(2J-2k+2)}
{(2k)(2J-2k+1)}
\right ) ~+~  (c-x^3) \\ \nn 
&=& - \frac{F(\xi,J)}{K J}  ~+~ (c-x_3) ~.
\eear
On the last line, we can make an approximation by adding an  exponentially
small `tail' to the sum, so that the function $F(\xi,J)$ 
will no longer depend on $J$, making $c-x_3$ be of order $J^{-1}$. 
Explicitly, we have
\ba
F(\xi,J) &:= c \sum_{m=0}^{\lfloor J \rfloor} ~2m
~ \left ( \left | \xi \right |^{2m}~
\prod_{k=1}^m \frac
{(2k-1)(2J-2k+2)}
{(2k)(2J-2k+1)}
\right )   \\ 
&\approx   c \sum_{m=1}^{\infty} ~2m
~ \left ( \left | \xi \right |^{2m}~
\frac {(2m-1)!!}{(2m)!!} 
\right ) 
 ~=~
 c |\xi| \frac{dK}{d|\xi|}~.
\end{align}
Taking the change in the eigenvalue to be zero, we get that
\be
c-x_3 ~=~  c J^{-1} \xi \frac{d (\ln K)}{d\xi} 
~=~ c J^{-1} \frac{|\xi|^2}{1-|\xi|^2} ~=~ 
J^{-1} c \frac{(c_1-c_4)^2 +c_2^2} {4c_1c_4 } ~.
\label{eq-correction-ellipsoid}
\ee
We have tested the correctness of this formula numerically,\footnote
{To facilitate numerical study, it is best to rewrite
equation (\ref{zero-eigenvector}) in as a genuine eigenvalue equation.  
Consider the operator $\sigma_3 \Heff$.  We can rewrite equation
(\ref{zero-eigenvector}) as 
$\left (-i\sigma_2\otimes (X_1-x_1) + i\sigma_1\otimes(X_2-x_2) + {\mathbf 1}_2\otimes X_3
\right ) |\Lambda\rangle = x_3 |\Lambda\rangle$.  Therefore, to find $x_3$ on
the emergent surface at a given $x_1$ and $x_2$, all we have to do is to solve
an eigenvalue problem.  It is important that the operator being diagonalized is
no longer hermitian: most (or possibly all) of its eigenvalues are complex.
Real eigenvalues (if any) correspond to points on the emergent surface.  Since
the dimension of the operator is even, there must be an even number of real eigenvalues
in non-degenerate cases.  This naturally corresponds to such points on the emergent
surface coming in pairs for a closed surface.}
as can be seen in figure \ref{f1}.

\begin{figure}
\includegraphics[width=6.5in]{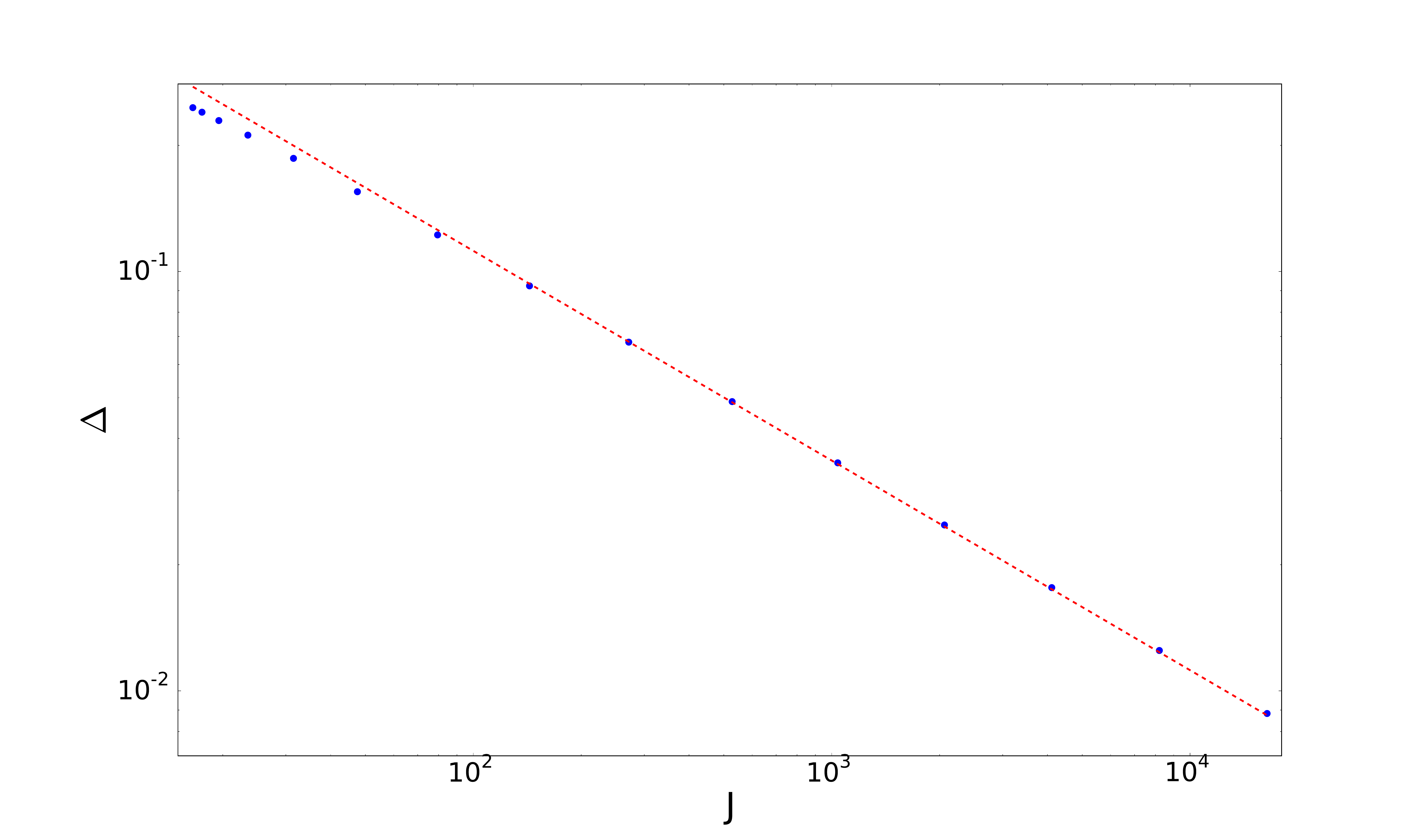}
\caption{
Magnitude, $\Delta$, of the difference between the approximate eigenvector 
and the exact eigenvector as obtained numerically, for the ellipsoid in figure 
\ref{f1}. The straight line, shown to guide the eye, is a best fit to the last few points and 
corresponds to $\Delta = \frac{1.12}{\sqrt{J}}$.}
\label{f2}
\end{figure}

Further, we have checked that
$\left [ \begin{array}{c} |\alpha\rangle \\ \hline 0 \end{array}\right ]$
is a good approximation to the exact eigenvector.  As can be
seen in figure \ref{f2}, the magnitude of the difference decreases as $N^{-1/2}$.

Once we understand $|\alpha\rangle$, we can ask about the leading correction
to the exact eigenvector of $\Heff$.  To next order, the eigenvector has a form
$\left [ \begin{array}{c} |\alpha\rangle + |\Delta\alpha\rangle \\ \hline |\beta\rangle \end{array}\right ]$, 
with corrections $|\beta\rangle$ and $|\Delta\alpha\rangle$ that have magnitudes of order no larger than $N^{-1/2}$.
Because we are working at a point where the normal vector points `up',
we have $\langle \alpha | \beta \rangle = 0$.  However,  generically
$\langle \Delta \alpha | \beta \rangle \neq 0$, so the actual
normal vector will show a small deviation from this assumed direction.
Finally,  $A_p \approx \sqrt{||\beta||^2-|\langle \Delta \alpha | \beta \rangle|^2}$.

It is difficult to obtain a closed-form formula for $|\beta\rangle$,
and even harder to obtain one for $|\Delta\alpha\rangle$.
We should proceed by finding a complete
eigenbasis for the first part of $\Heff$ as written in equation
(\ref{heff-two-parts-ellipsoid}), and then use standard 
perturbation theory to obtain the desired result.  This is
beyond the scope of this paper, so we will resort to
less complete methods to obtain some insight into the structure.

The formal expression for $|\beta\rangle$ is
\be
 |\beta\rangle ~=~
(A^\dagger)^{-1}~ \left ( c\frac{L_3}{J}  - x_3 ~+~p^{(2)}_3 \right ) |\alpha\rangle~.
\label{beta-formal}
\ee
This expression is formal because $A^{\dagger}$ might not have
an inverse when acting on the above operator.  However, we notice that
since we already know $x_3$, we are able to find,
to leading order in $N$, the first
nonzero coefficient of $|\beta\rangle$ (which is the coefficient of $|J-1\rangle$).  To do so,
we take our already computed value of $x_3$ and solve this equation:
\be
A^\dagger |\beta \rangle = - \left ( c\frac{L_3}{J} - x_3 \right) |\alpha \rangle~.
\ee
Once we have the first coefficient, we can substitute it back into the 
above equation and solve for the next coefficient.  Repeating
this will in principle yield nearly all components of $|\beta\rangle$ (with exception of the
component with the most negative $L_3$ eigenvalue).

Explicitly, we obtain that the coefficient of
$|J-1\rangle$ in $|\beta\rangle$ is
\be
\frac{c-x_3}{\sqrt K}\frac{\sqrt{2J}}{c_1-c_4-ic_2}~.
\ee

The magnitude squared of this expression is
\be
\frac{c^2}{2J}\frac{1}{\det{\mathbf C}}\frac{|\xi|^2}{(1-|\xi|^2)^{1/2}}~.
\label{eq-leading-coeff-of-beta}
\ee
We need this expression to be small (compared to 1), since
we would like $\|\beta\| \ll \|\alpha\|$. Thus, for nonzero $|\xi|$, 
how large $J$ needs to be for our analysis 
to be applicable depends, for example, on $c$.  
Numerical study confirms equation  (\ref{eq-leading-coeff-of-beta}); further,
it shows that the ratio of the expression in equation (\ref{eq-leading-coeff-of-beta})
and the total magnitude squared of $|\beta\rangle$ goes to a constant value at
large $N$.  Thus, $\|\beta\|^2$ is proportional to $c^2$ and decreases with large $J$ 
like $J^{-1}$.

We will see in section \ref{largeN} that corrections shown in equations
(\ref{eq-correction-ellipsoid}) and (\ref{eq-leading-coeff-of-beta}) are large
when $N$ is too small to describe the portion of a given surface with a high curvature.

At the same order, we also get a correction to $|\alpha\rangle$,
$|\Delta \alpha\rangle$.  A formal expression, similar to the one for $|\beta\rangle$ above,
\be
|\Delta \alpha\rangle ~=~
A^{-1} \left ( (a+ib)\left(\frac{L_3}{J} -1\right)  \right )
|\alpha\rangle~,
\label{alpha-formal}
\ee
does not have a well defined meaning as 
$ \left ( (a+ib)\left(\frac{L_3}{J} -1\right) \right ) |\alpha\rangle$
generically has a significant component parallel to $|\alpha\rangle$.
It is not possible to solve for coefficients of $|\Delta \alpha\rangle$
in the same way that we solved for those of $|\beta\rangle$;
we need a complete perturbation theory treatment.  However,
using the above expression as a guide to structure at least, we see that
the correction $|\Delta \alpha\rangle $
is of order ${\cal O}(N^{-1/2})$, and that it would grow with
$a$ and $b$.  While the coefficient $c$ determines the 
local curvature of the surface, the coefficients $a$ and $b$
control how fast the noncommutativity is changing, as we will see
in section \ref{subsection:nc}. 

As we already mentioned, $|\Delta \alpha\rangle$
is not necessarily orthogonal to $|\beta\rangle$, so we will 
now have a correction to the angle of the normal vector,
\be
n_i \approx (2\Re \langle\Delta \alpha|\beta\rangle, 2\Im \langle\Delta \alpha|\beta\rangle, 1)~.
\label{normal-vector-wobble}
\ee
Numerical work confirms that the angle between
the expected normal vector to the surface (which here points
in the $x_3$-direction) and the actual normal vector to the
surface scales like $N^{-1}$ and grows linearly with
the coefficients $a$ and $b$.  
We will return to this point in section \ref{largeN}.

\subsection{Polynomial maps from the sphere}
\label{subsection:generic}

\begin{figure}
\includegraphics[width=6.5in]{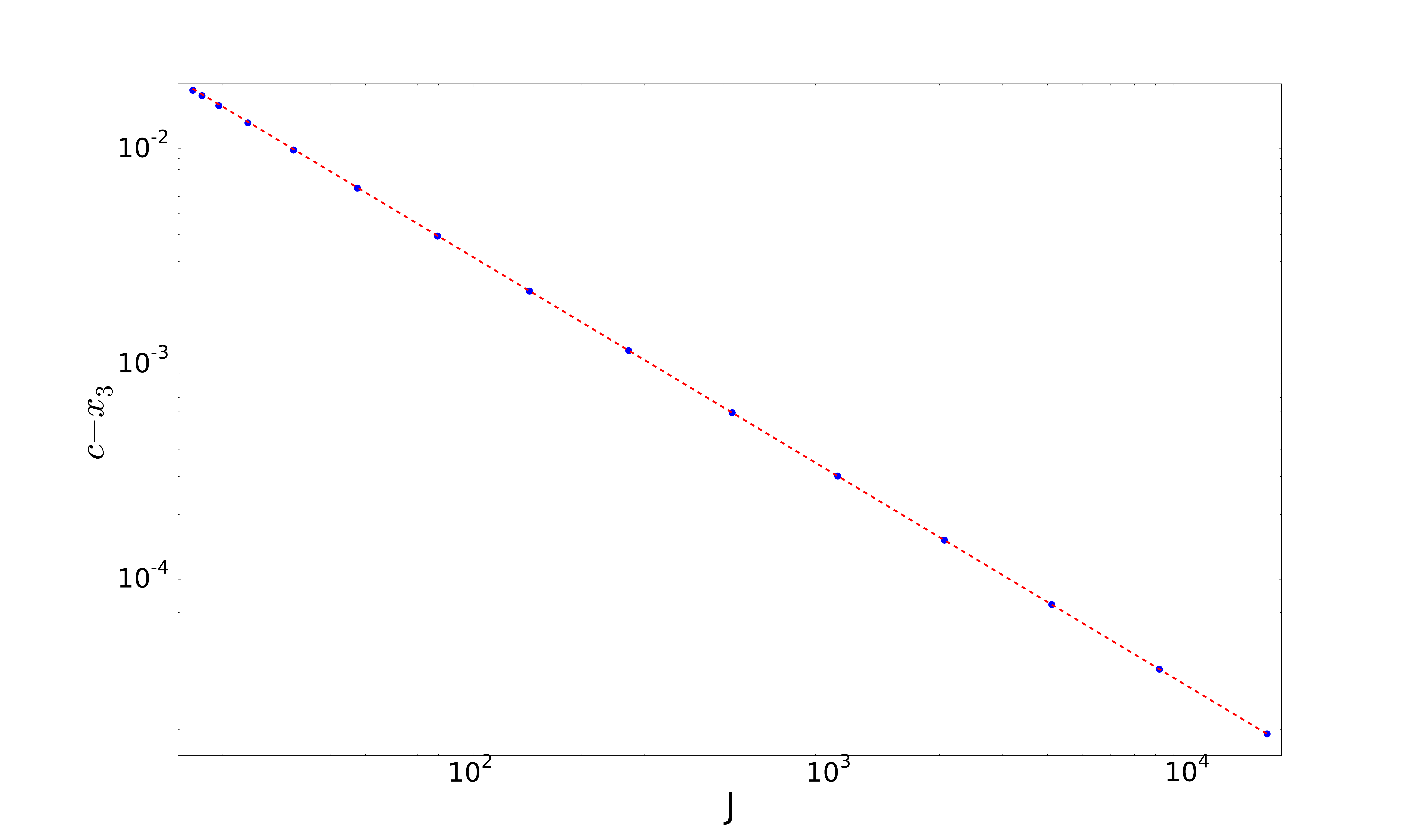}
\caption{The difference between the actual eigenvalue $x_3$ and the classical (large $N$)
position $c$ for a generic surface given by
$x_1=1+w_1+0.5w_3$, $x_2=2w_2$, $x_3=w_3+0.2 w_1 w_2$, 
at a point given by $(w_1,w_2,w_3)=(1/2,1/4,\sqrt{11}/4)$.
The line shows equation (\ref{general-correction}).
}
\label{f3}
\end{figure}

Our analysis of a generic polynomial surface will build on the
analysis of an ellipsoid.  
Consider a point of interest such that
the normal at this point is pointing in the positive $x_3$ direction.
Let this point lie at $x_1=x_2=0$, setting $y_1=y_2=0$.  Without loss
of generality, set $y_3$ equal to zero as well.
This allows us to write $\Heff$ as a sum of two pieces as before:
\bear
\Heff(x_3) &=& \left [ \begin{array}{c|c}
 0 & A^\dagger
\\ \hline 
A & 0
\end{array} \right ] \label{heff-two-parts-generic}\\ 
\nn &+&
\left [ \begin{array}{c|c}
c\frac{L_3}{J}  - x_3 ~+~p^{(2)}_3 & 
(a-ib)\left(\frac{L_3}{J} -1\right) ~+~ p^{(2)}_1 - ip^{(2)}_2
\\ \hline 
(a+ib)\left(\frac{L_3}{J} -1\right) ~+~ p^{(2)}_1 + ip^{(2)}_2
  &-\left(c\frac{L_3}{J} -x_3\right)~-~p^{(2)}_3
\end{array} \right ]~.
\eear
$p^{(2)}$ are the polynomials introduced in section \ref{subsection:variety}:
to leading order, they can be written as
\bear
p^{(2)}_k &=& d_{k,1}\left (\frac{L_+}{J}\right)^2 + 
d_{k,2}\left (\frac{L_-}{J}\right)^2 +
d_{k,3}\,\frac{L_+L_-+L_-L_+}{2J^2} \\
&=& e_{k,1}\left(\frac{L_1}{J}\right)^2 +
e_{k,2}\left(\frac{L_2}{J}\right)^2 +
e_{k,3}\,\frac{L_1L_2+L_2L_1}{2J^2} 
\label{poly-leading}
\eear
where $e_{k,1} = d_{k,1} + d_{k,2} + d_{k,3}$,
$e_{k,2} = -d_{k,1} - d_{k,2} + d_{k,3}$ and 
$e_{k,3} = 2i(d_{k,1} - d_{k,2})$.
Second or higher order polynomials containing at least one power of $L_3/J-1$ are
either equivalent to polynomials in $L_1/J$ and $L_2/J$ (from $L_1^2+L_2^2+L_3^2=N^2-1$),
or subleading, as we will see in a moment.

The vector defined in equation (\ref{eigenvector-leading-order}) 
together with  $|\alpha \rangle$  given in equation (\ref{alpha}) is an approximate zero 
eigenvector of this more general $\Heff$ as well,
as we have confirmed numerically. 
Generically, $A_p$ decreases with large $N$ like $N^{-1/2}$.

Analytically, we first compute the following quantities
\ba
 \langle \alpha|\frac{L_{-}L_{+}}{J^2}|\alpha \rangle &\approx \frac{2}{J}\frac{|\xi|^{2}}{1-|\xi|^2} \\
 \langle \alpha|\frac{L_{+}L_{-}}{J^2}|\alpha \rangle &\approx \frac{2}{J}\frac{1}{1-|\xi|^2} \\
 \langle \alpha|\frac{L_{+}L_{+}}{J^2}|\alpha \rangle &\approx \frac{2}{J}\frac{\xi}{1-|\xi|^2} \\
 \langle \alpha|\frac{L_{-}L_{-}}{J^2}|\alpha \rangle &\approx \frac{2}{J}\frac{\bar \xi}{1-|\xi|^2} 
~.
\end{align}
These imply that corrections to $x_3$ due to the polynomials
$p^{(2)}_k$ in equation (\ref{poly-leading}) are of order $J^{-1}$, same as
correction in equation (\ref{eq-correction-ellipsoid}).  In fact, we can
compute the new  corrections to the eigenvalue $x_3$ in this case:
\be
c-x_3 = \frac{1}{J} ~\left (
c\,\frac{|\xi|^2}{1-|\xi|^2}~-~\frac{
|1+\xi|^2 e_{3,1} + |1-\xi|^2 e_{3,2} + i(\xi-\bar\xi) e_{3,3}}{2(1-|\xi|^2)}
\right )~.
\label{general-correction}
\ee

Figure \ref{f3} shows comparison between this approximate result and the exact numerical values.
The agreement is excellent.

To summarize the size of the various
higher order corrections, we notice that
\bear
\left \| ~(L_3/J-1)|\alpha\rangle ~\right \| &\sim& {\cal O}(N^{-1})
\label{L3-vanishes}
\\
\left \| ~(L_1/J)|\alpha\rangle ~\right \| &\sim& {\cal O}(N^{-1/2})
~~~\mathrm{and}~~
\left \| ~(L_1/J)^2|\alpha\rangle ~\right \| ~\sim~ {\cal O}(N^{-1})
\label{L1-vanishes}
\\
\left \| ~(L_2/J) |\alpha\rangle ~\right \| &\sim& {\cal O}(N^{-1/2})
~~~\mathrm{and}~~
\left \| ~(L_2/J)^2|\alpha\rangle ~\right \| ~\sim~ {\cal O}(N^{-1})~.
\label{L2-vanishes}
\eear

To go further in our analysis, we could ask how introducing higher-order
polynomials affects $|\beta\rangle$ and $|\Delta\alpha\rangle$ 
(and therefore $A_p$ as well as the angle the actual normal vector
makes with its expected direction), 
or more generally, what is the effect of all these terms on the
exact eigenvector.  
The analysis parallels one at the end of the previous subsection:
coefficients of the quadratic terms in $p_3^{(3)}$ enter in the 
same way that $c$ does and 
coefficients of the quadratic terms in $p_3^{(2)}$
and $p_3^{(2)}$ enter in the same way that $a$ and $b$ do.
Thus, again, having a larger curvature on the surface affects 
$\|\beta\|^2$ while having the noncommutativity vary quickly
affects $\|\Delta\alpha\|^2$ (as we will see).

As before, formulas for the first few coefficients of $|\beta\rangle$ can be computed
recursively.  The results are too complicated to be illustrative, however, they
are qualitatively similar  to those for the ellipsoid: $\|\beta\|^2$ falls off
like $1/J$, grows with $c^2$ and quadratically with the coefficients in $p^{(2)}_3$
and depends in a nontrivial way on $|\xi|$.  In contrast to the ellipsoid
case, it is possible for $\|\beta\|^2$ to be nonzero even with zero $|\xi|$.

Finally, even higher order polynomials are proportionately more
suppressed.  For example terms involving $(L_3/J-1)^2$ are suppressed by
$N^{-2}$:
\be
 \langle \alpha|\left(L_3/J - 1\right)^2|\alpha \rangle ~\approx~ \frac{1}{J^2}\frac{|\xi|^2(2+|\xi|^2)}{(1-|\xi|^2)^2} ~.
\ee

\subsection{Local noncommutativity}
\label{subsection:nc}

Consider $-i[X_1,X_2]$, using the form in equation (\ref{polynomial}).  We have
\bear
-i[X_1,X_2] = (c_1 c_4 - c_2c_3) (L_3 /J^2)&+&\mathrm{terms~linear~in~}(L_1/J^2) 
\mathrm{~and~}(L_2/J^2)\nn \\
&+&\mathrm{terms~with~higher~powers~of~}L_i~. 
\eear
From the formulas in section \ref{subsection:generic}, the expectation value of this
operator in the coherent state is just
\be
\theta_{12}~=~\langle \alpha | -i[X_1,X_2] |\alpha \rangle = (c_1 c_4 - c_2c_3)/J~,
\label{nc-from-C}
\ee
since the corrections to $\langle \alpha | L_3/J |\alpha \rangle \approx 1$, as well as
 $\langle \alpha | L_1/J^2 |\alpha \rangle$, $\langle \alpha | L_2/J^2 |\alpha \rangle$
and those terms that are higher order (in $L_i$s), all lead to subleading contributions (of 
order $1/J^2$ or smaller).
It is important to insist that $c_1 c_4 - c_2 c_3$ is nonzero,
so the leading contribution above does not vanish.

\begin{figure}
\begin{center}
\includegraphics[width=6.5in]{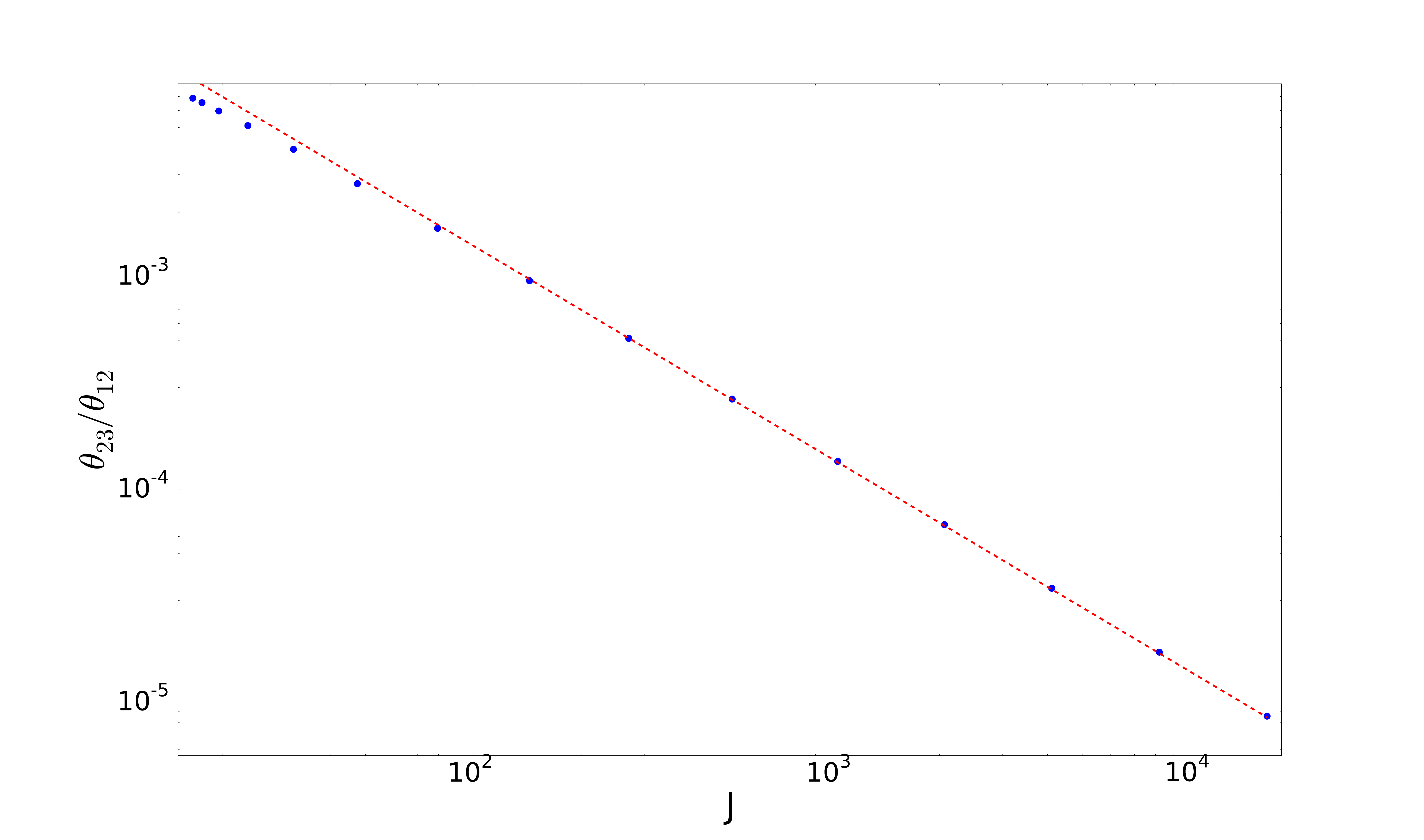}
\end{center}
\caption{$\theta_{23}/\theta_{12}$ for the example in figure \ref{f3}.
This ratio appears to decrease like $J^{-1}$.}
\label{f4}
\end{figure}

We can examine $\langle \alpha | -i[X_1,X_3] |\alpha \rangle$ and  $\langle \alpha | -i[X_2,X_2] 
|\alpha \rangle$
in a similar way.  In this case, only  those sub-leading terms are nonzero and we obtain that
\be
\theta_{i3}~=~ \langle \alpha | -i[X_i,X_3] |\alpha \rangle~\sim~ 1/J^2~~\mathrm{for}~i=1,2.
\ee
Therefore, we have that $\theta_{i3}/\theta_{12}$ is of order $1/J$, which is well supported
by our numerical data (see Figure \ref{f4}).   We can then take $\theta = \theta_{12}$.
A more general, rotationally invariant equation is
\be
\theta = \langle \alpha| ~
\Theta~|\alpha \rangle~~,~~~
\mathrm{where}~\Theta := \sqrt{-\sum_{i\neq j}[X_i,X_j]^2}~.
\label{theta-general}
\ee
We have introduced a new operator, $\Theta$, which will play an important
role in the next section.

Equation  (\ref{nc-from-C}) has a simple geometric interpretation: the
local noncommutativity on the round sphere is constant and equal to $1/J$.
A single noncommutative `cell' with this area is mapped to an ellipse
with area $(\det\,\boldsymbol{C})/J$, which is just the noncommutativity in
equation  (\ref{nc-from-C}).  In other words, the local noncommutativity
is the volume form on the emergent surface divided by the volume form
on the sphere, times $J^{-1}$.

The local noncommutativity is not constant on the surface.  An explicit
computation on the ellipsoid in equation (\ref{ellipsoid-def}) shows that its derivatives are 
\begin{align}
\frac{\partial \theta}{\partial x}&=\frac{b(c_1 c_3+c_2c_4)-a(c_3^2+c_4^2)}{(c_1c_4-c_2c_3)J} ~~~\mathrm{and} 
\label{diff-theta-x}\\
\frac{\partial \theta}{\partial y}&=\frac{a(c_1 c_3+c_2c_4)-b(c_1^2+c_2^2)}{(c_1c_4-c_2c_3)J}~.
\label{diff-theta-y}
\end{align}
If we include higher order polynomials, the appropriate coefficients in $p_1^{(2)}$
and $p_2^{(2)}$ enter in the same way as $a$ and $b$ above.  Thus, we see
that having these coefficients larger makes the noncommutativity
vary faster, as we have mentioned before.

\subsection{Coherent states overlaps, U(1) connection and $F_{\mu\nu}$ on a D2-brane}
\label{subsection:coherent}

Since coherent states are associated with points, it is important that
the overlap between coherent states corresponding to well-separated points
be small.
\begin{figure}
\includegraphics[width=6.5in]{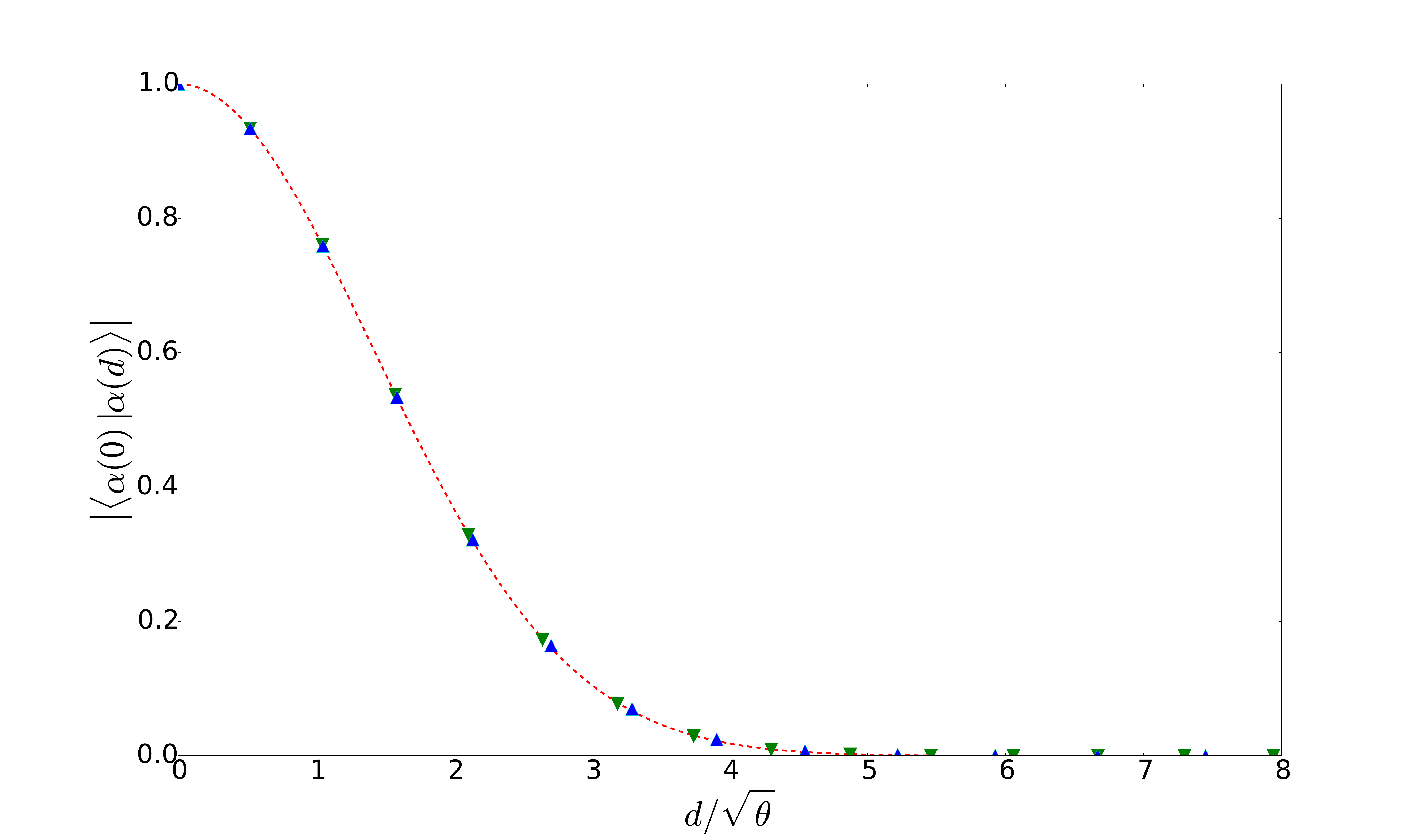}
\caption{Magnitude of the overlap between the eigenstate corresponding to
the point p at the north pole and the eigenstate corresponding to a point
p' a distance $|d|$ away. The green
$\blacktriangledown$ corresponds to points p' with $x_2=0$, while
the blue $\blacktriangle$ corresponds to p' with $x_1=0$.
The dashed line corresponds to equation (\ref{overlap}). 
Plotted for  an ellipsoid with $c_1$=1, $c_2$=0.75, $c$=12, with $N$=16,384.
}
\label{f6}
\end{figure}
 Consider two points p and p$'$ on the emergent surface which
are within a distance of order $1/\sqrt N$ of each other.  For large $N$,\footnote
{The question of what constitutes a large enough $N$ is discussed in
section \ref{largeN}.} the coefficients $c_i$, $a$, $b$, $c$ etc$\ldots$
that locally characterize the surface are approximately the same.
However, the corresponding pre-images of p and p$'$ on the unit sphere
in $w$-space are sufficiently far apart that the basis in which
equation (\ref{alpha}) is written is completely different.
Therefore, the approximate coherent state at the point p$'$ can be obtained
from the coherent state at the point p by an SU(2) rotation (in
the $N$-dimensional representation).  Explicitly,
\be
|\alpha'\rangle = e^{i(-D_2 L_1 + D_1 L_2)}~|\alpha\rangle ~,
\ee
where $D_1$ and $D_2$ are small displacements in $w$-space corresponding
to moving from p to p$'$. Since we have positioned p at the north pole
of the unit sphere, there is no displacement in the 3-direction.
$L_1$ and $L_2$ can be written in terms of $A$ and $A^\dagger$
via equation (\ref{A}), and we get that
\be
|\alpha'\rangle = e^{\frac{i}{2\theta}(d A + \bar d A^\dagger)}~|\alpha\rangle ~,
\label{alphap}
\ee
where $d=x'_2-ix'_1$, with $x'_1$, $x'_2$ being the coordinates of point p$'$.
To compute the overlap between $|\alpha\rangle$
and $|\alpha'\rangle$, we use the Baker-Campbell-Hausdorff formula to 
leading order, together with $A|\alpha\rangle \approx 0$:
\be
\langle \alpha |\alpha'\rangle ~=~ 
\langle \alpha |~e^{-\frac{1}{8\theta^2} d\bar d[A,A^{\dagger}]}~|\alpha\rangle 
~\approx~ e^{-\frac{|d|^2}{4\theta}} ~,
\label{overlap}
\ee
since $[A,A^{\dagger}] = 2 \theta (L_3/J)$.  As can be seen in figure \ref{f6},
the actual coherent states have exactly this expected behaviour.

Further, we can look at the connection defined (to within a factor of 2)
in equation (28) of \cite{Berenstein:2012ts},
\be
2 v^iA_i = -iv^i\langle \alpha(x_i) |\partial_i|\alpha(x_i)\rangle ~,
\ee
where $v^i$ is a tangent vector on the emergent surface.  To evaluate it,
we rewrite equation (\ref{alphap}) in terms of the small displacements 
$x_1$ and $x_2$:
\be
|\alpha'\rangle = e^{\frac{i}{\theta}(-x_2 X_1 + x_1 X_2)}~|\alpha\rangle ~.
\ee
Thus, the connection is just 
$(A_1,A_2) = (-x_1/2\theta,x_2/2\theta)$ and the curvature is $F_{12}=\theta^{-1}$.
This is exactly the expected result on an emergent D2-brane \cite{Seiberg:1999vs}.

\subsection{Nonpolynomial surfaces}
\label{subsection:exotics}

Not surprisingly, our general conclusions are applicable even
when the maps from the sphere to the surface of interest are not
polynomial.  As long as the maps are smooth enough to be 
approximated by a Taylor polynomial, the large $N$ limit
behaviours should be similar.  Examples with many
desired properties can be relatively easily `cooked up'.  Here
we consider two of conceptual relevance.

Our first example using a non-polynomial map is designed to 
probe into the role of the parameter $\xi$.  To this end, we 
examine
\ba
x_1 &= w_3 w_1 + \sqrt{1-w_3^2} ~w_2 \label{eq1}~,\\
x_2 &= -\sqrt{1-w_3^2}~ w_1 +w_3 w_2 \label{eq2}~.\\
x_3 &= w_3~.  \label{eq3}
\end{align}

This example is designed produce a round sphere 
with a constant local noncommutativity $\theta$ by
`shearing' the original sphere (to preserve the volume form).  We have
checked explicitly that $\theta$ is constant over the surface and equal to $1/J$
in the large $N$ limit.  The parameter $\xi$, however, is not
constant, instead, we have $\xi =-i\sin(\phi)/(2-i\sin(\phi))$.
This shows that $\xi$ does not play a role in the large $N$ limit of
the surface: it can be changed by applying a volume preserving automorphism to the
sphere. Another way to look at it is that the three matrices $X_i$ defined
by equations (\ref{eq1})-(\ref{eq3}) can be obtained from $L_i/J$ by
a conjugation (up to some ordering ambiguities).  $\xi$ can thus
be viewed as a basis-dependent quantity.

\begin{figure}
\includegraphics[width=6.5in]{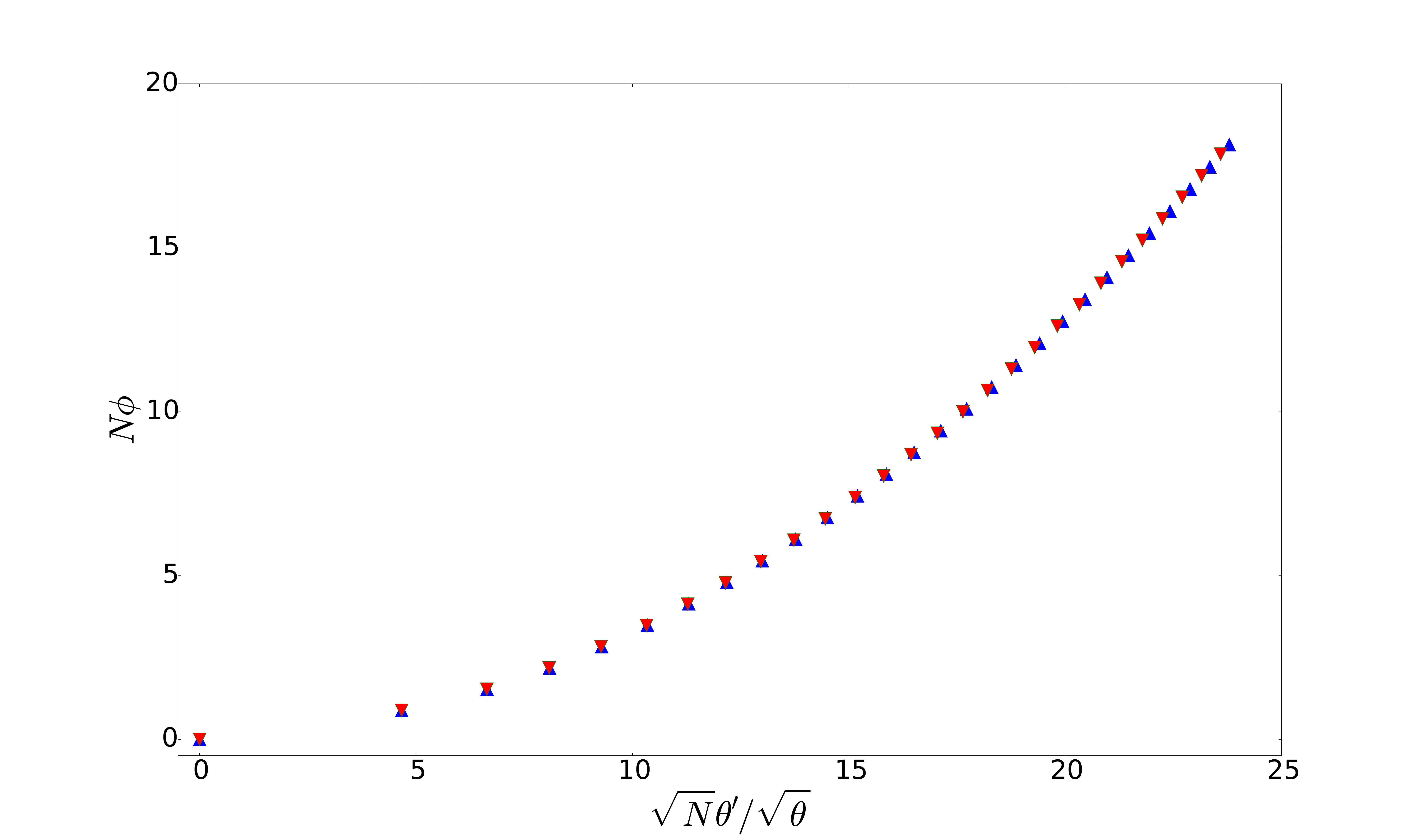}
\caption{
Angle $\phi$ between the normal vector $\vec{n}$ computed using equation 
(\ref{normal}) and the noncommutativity vector
$\epsilon_{ijk}\theta_{jk}$,
for the surface in equation (\ref{example2}) at a point given by
$x=0.5$, $y=0$. The blue $\blacktriangle$ corresponds to N=3000 and the red
$\blacktriangledown$ to N=12 000; the agreement between plots at different
$N$ shows that the plotted quantities scale with $N$ in the expected way.
On the horizontal axis we have a derivative of the noncommutativity
along the surface scaled by $\sqrt{\theta}$, which increases
as $\mu$ is increased in equation (\ref{example2}).
}
\label{f7}
\end{figure}

Another interesting example is given by
\ba
x_1 &= \frac{w_1}{\sqrt{w_1^2+ w_2^2 + \mu^2 w_3^2}}~, \nn\\
x_2 &= \frac{ w_2}{\sqrt{w_1^2+ w_2^2 + \mu^2 w_3^2}} ~,
\label{example2}
\\
x_3 &= \frac{\mu w_3}{\sqrt{w_1^2+ w_2^2 + \mu^2 w_3^2}}~. \nn
\end{align}
In this example, we again get a round sphere, but the local noncommutativity
is no longer constant.   As we would expect, the actual surface at
finite $N$ differs from a round sphere at order $1/N$; this corresponds to
the normal vector deviating from the radial direction at the same order,
as given by equation (\ref{normal-vector-wobble}).
Further, we can compute the noncommutativity vector $\epsilon_{ijk}\theta_{jk}$.
Our assertion is that these two vectors should be nearly parallel.  Figure
\ref{f7} shows that, indeed, the angle between these two vectors
decreases as $1/N$.  This angle increases as the coefficient
$\mu$ is increased, resulting in a more rapidly changing noncommutativity.
Interestingly, $A_p$ turns out to be subleading, of order
$1/N^{3/2}$ or smaller, instead of $1/N^{1/2}$, implying that $|\Delta \alpha\rangle$
is nearly parallel to $|\beta\rangle$.

The two examples in this subsection demonstrate that our approach works for
surfaces which are not given by polynomial maps from the sphere.  This
is not surprising, as our approach should work for any surface which can be 
locally approximated by a polynomial map over the sphere.  Relaxing the 
polynomial condition allows for just about any smooth surface which is topologically
equivalent to a sphere to be studied with our approach.

\section{Large $N$ limit and the Poisson bracket}
\label{largeN}

In the previous section, we have provided a series of examples
increasing in generality and all sharing the following common features:
there existed a family of matrix triplets $X_i$ labeled by their 
size $N$.  Each such triplet give rise to a surface $\mathcal{S}_N$ given by the
locus of points where $\Heff(x_i)$ had a zero eigenvalue.  
The zero eigenvector of $\Heff$ at a point on a surface such
that the normal to this surface was pointing in the $x_3$
direction was, either exactly or approximately, of the form
\be
\left [ \begin{array}{c} |\alpha\rangle \\ \hline 0 \end{array}\right ]~.
\ee
Where the zero eigenvector was not exactly of this form, the corrections
were small, of order $N^{-1/2}$.

More generally, since a rotation
of the coordinate system can be effected by an SU(2) rotation of 
the $\sigma_i$ matrices in $\Heff$, the zero eigenvector at
an arbitrary point $p$ has the form
\be
|\Lambda_p \rangle~=~ \left [ \begin{array}{c}  |\alpha_1\rangle \\ \hline  |\alpha_2\rangle  \end{array}\right ]~=~
 \left [ \begin{array}{c} a |\alpha_p\rangle \\ \hline b |\alpha_p\rangle  \end{array}\right ]
~+~{\cal{O}}\left ( N^{-1/2} \right )
\label{parallel}
\ee
where $|a|^2 + |b|^2 = 1$ and where $|\alpha_p\rangle$ is a unit $N$-dimensional vector.

Given the two parts of a zero eigenvector of $\Heff$, $|\alpha_1\rangle$
and $|\alpha_2\rangle$, at finite $N$, we compute $|\alpha_p\rangle$ as follows:
find the normal vector to the surface, $n_i = \langle \Lambda_p |\sigma_i |\Lambda_p \rangle$.  Then, find the SU(2) rotation that
brings this vector to point in the positive $x_3$ direction and 
apply it to $\Lambda_p$.  Then, the top component of of $|\Lambda_p\rangle$ is $|\alpha_p\rangle$.
Explicitly,
\be
|\alpha_p\rangle ~~=~~\cos(\theta_{\hat n}/2) e^{i\phi_{\hat n}/2} |\alpha_1\rangle~+~
\sin(\theta_{\hat n}/2) e^{-i\phi_{\hat n}/2}|\alpha_2\rangle~,
\ee
where $\theta_{\hat n}$ and $\phi_{\hat n}$ are the polar angles of the 
unit normal vector $\hat n$.

Once the coherent state $|\alpha_p\rangle$ corresponding to a point is identified,
we can define a correspondence between functions
on the large-N surface $f$ and operators ($N\times N$ matrices) $M_f$
through
\be
f(\tau) = \langle \alpha_p| M_f |\alpha_p\rangle~,
\ee
where $\tau=(\tau_1,\tau_2)$ is a coordinate of some point $p$ on the surface.

The function $s: M_f \rightarrow f$ is usually called the symbol map; using
a coherent state to define the symbol is an approach due
to Berezin \cite{Berezin:1974du}.  
The implied noncommutative star product is 
\be
(f \star g)(\tau) := 
\langle \alpha_p| M_f \, M_g |\alpha_p\rangle~.
\ee

The star product is not unique, ie it is not fixed by the surface and the 
noncommutativity parameter $\theta$ alone.  There are many different 
triplets of matrices that give the same surface and noncommutativity; 
different triplets would lead to different star products. Only the leading 
order of the commutator $f \star g - g \star f \approx \theta $ is 
universal. For example, the details of the star product depend on $\xi$ 
which we know to be arbitrary.  However, the star product implies,
in the large $N$ limit, a unique antisymmetric bracket,
\be
\{f,g\} := N \left (f \star g - g \star f \right )~.
\ee
We would like this bracket to give us a Poisson structure
on our emergent surface.  It is naturally skew-symmetric and satisfies
the Jacobi identity, so it is a Lie bracket.  To be a Poisson
bracket, it also needs to satisfy the Leibniz Rule:
\be
\{fg,h\} = f\{g,h\} + g\{f,h\}~.
\ee
(Notice that these are ordinary multiplications now, not
star-products.)

Instead of directly proving that the Leibniz Rule holds, 
we will show that our definition of a star product is equivalent
to
\be
\{f,g\} = \frac{1}{\rho} \epsilon^{ab} \, \partial_a f \, \partial_b g
\label{bracket}
\ee
for some function $\rho$ on the surface.  In particular, we will have
\be
\rho = \frac{\sqrt{\det g}}{N\theta}~,
\label{rho}
\ee
where $g$ is the pullback metric on the noncommutative surface 
and $\theta$ is the local noncommutativity parameter defined in
subsection \ref{subsection:nc}.

Let's follow our previous approach, and consider not only
$X_i$ to be polynomials in $L_1/J$, $L_2/J$ and $L_3/J-1$, but also
consider operators that are polynomials
in $X_i$ (and therefore polynomials in $L_1/J$, $L_2/J$ and $L_3/J-1$).
The degrees and coefficients of all the  polynomials are fixed while
$N\rightarrow \infty$.  First, consider the expectation value 
$\langle \alpha_p| \, M \,|\alpha_p\rangle$ 
of some such operator $M=m(X_1,X_2,X_3)$ in a coherent state,
where $m(\cdot,\cdot,\cdot)$ is a polynomial function.
We can compute $\langle \alpha_p | \, M \,|\alpha_p\rangle$
at a point $p$ where the normal points straight up 
by first writing $M$ as a polynomial in $L_1/J$, $L_2/J$,  and $(L_3-1)/J$.
Then, from equations (\ref{L3-vanishes}), (\ref{L1-vanishes})
and (\ref{L2-vanishes}), we see that the leading order piece (which
stays finite as $N \rightarrow \infty$) is simply the constant term\footnote
{Any ambiguities due to the fact that $L_1^2 + L_2^2 + L_3^2 = N^2-1$
are subleading in N.}.   Thus, 
\be
\langle \alpha_p | \, M \,|\alpha_p\rangle = m(y_1,y_2,y_3)~,
\ee
where $y_i$ are the coordinates of the surface at point $p$  as
 defined in equations (\ref{polynomial}).

Now that we have shown that the expectation value in a coherent state at a point
of any polynomial (in $X_i$) operator is exactly what we would expect, let's
think about the expectation value of the commutator of two such operators
$M_1$ and $M_2$.  Consider then two polynomials, $m_1$ and $m_2$ in $x_1$, $x_2$ and $x_3$,
and the corresponding operators
$M_1=m_1(X_1,X_2,X_3)$ and $M_2=m_2(X_1,X_2,X_3)$.
We have already argued that $\theta_{12}$ is much larger
than $\theta_{13}$ and $\theta_{23}$.  A similar argument extended to functions
of $X_i$ shows that, as long as $X_i$s are of the form (\ref{polynomial}), we have
\be
\langle \alpha_p | \, -i[M_1,M_2] \, | \alpha_p \rangle~=~
\theta_{12} \left (
\frac{\partial m_1(y_1,y_2,y_3)}{\partial y_1} \frac{\partial  m_2(y_1,y_2,y_3)}{\partial y_2} 
-
\frac{\partial m_1(y_1,y_2,y_3)}{\partial y_2} \frac{\partial  m_2(y_1,y_2,y_3)}{\partial y_1}  
\right )~.
\ee
Thus, for the two functions on the noncommutative surface given as
restrictions of the polynomials $m_a$: $f_a(\sigma) = m_a(x_i(\sigma))$, the
bracket is
\ba
\{f_1,f_2\} &= N \langle \alpha_p | \, [M_1,M_2] \, | \alpha_p \rangle  \\ \nn &= N\theta~
\left (\frac{\partial \sigma_a}{\partial x_1} \frac{\partial \sigma_b} {\partial x_2}
-
\frac{\partial \sigma_a}{\partial x_2} \frac{\partial \sigma_b} {\partial x_1}
\right )  
\frac{\partial f_1}{\partial \sigma_a} \frac{\partial f_2}{\partial \sigma_b}
= N\theta~\frac{\epsilon_{ab}}{\sqrt{\det g}}
\frac{\partial f_1}{\partial \sigma_a} \frac{\partial f_2}{\partial \sigma_b}~,
\end{align}
in agreement with equations (\ref{bracket}) and (\ref{rho}).

To summarize, we have proven that our emergent surface is equipped with 
natural Poisson bracket which satisfies the correspondence principle
\be
\{\cdot,\cdot\} ~~\leftrightarrow ~~-iN[\cdot,\cdot]~.
\ee

Essential for our argument to work was the noncommutativity
vector $\epsilon_{ijk} \theta_{jk}$ being nearly parallel to
the normal vector $n_i$, as shown in Figure \ref{f7}.  If this was not the case, the
bracket we defined would fail to be a Poisson bracket.

For the remainder of this section, we will answer the following question:
given a surface embedded in three dimensions and a Poisson structure
on this surface, does there exist a matrix description that approximates this surface?

Our construction gives a positive answer to this question, and provides 
restrictions on the surface and on $N\theta$ for the approximation to be 
good.  We focus on $N\theta$ (rather than $\theta$ itself) as this is a  finite quantity in the large
$N$ limit and determines the Poisson structure through equation (\ref{rho}).
Given a surface and a function $N\theta$ on this surface,
we can always define a map from the unit sphere to this 
surface such that the ratio of the volume form on the surface to the 
volume form on the sphere is $N\theta$ (see equation (\ref{c-matrix-metric})).  In fact, we can 
find many such functions.  Which we pick will affect $\xi$ and the higher 
orders of the star product, but not the overall noncommutative structure.
Note, however, that it is not possible to set $\xi$ to
zero everywhere for a generic noncommutative surface.  
$\xi$ is zero if the metric on the emergent surface is proportional
to the metric on the sphere, while the coefficient of this proportionality
must be the noncommutativity $\theta$, which is fixed.  These two 
requirements would fix (up to diffeomormisms) the metric on the emergent 
surface, which is already fixed by the embedding.  To view this in a different
way, the
freedom in choosing a map from the sphere to the emergent surface
is the freedom to pick two functions on the sphere.  One of these
functions is fixed by requiring a particular noncommutativity $\theta$.
The remaining function can be used to change $\xi$.  However, $\xi$ is
a complex function, so requiring it to vanish over-constrains the problem.

Given a map from the sphere to the desired surface, we need only
replace the rectilinear coordinates on the sphere with some SU(2) 
generators $L_i$ and we obtain a triplet of matrices $X_i$ which lead us 
to the appropriate noncommutative structure.
Here, again, there is ambiguity in the ordering of the operators.
Its effects are  suppressed by powers of $1/N$ and it affects
higher order terms in the star product (but not the leading order term).

For this construction to work, the surface we start with must be 
sufficiently smooth.  Alternatively, we could say that we need to pick
an irrep of SU(2) large enough to accommodate a rapidly varying surface.
Two conditions seem necessary: that the curvature radii of the surface at any 
point be much larger than the diameter of a noncommutative `cell'
($R_{\mathrm{curvature}} \gg \sqrt{\theta}~\sim~N^{-1/2}$) and that 
$\theta$ change slowly.  Let $\theta'$ be a derivative of $\theta$ in some tangent direction.
Then, the change in noncommutativity over a single cell (which has
an approximate diameter of $\sqrt{\theta}$), $\sqrt{\theta}\theta'$, should be
be small when compared with $\theta$ itself: $\theta'/\sqrt{\theta} \ll 1$ ($\theta'/\sqrt{\theta} \sim N^{-1/2}$).
As we have already discussed, 
in equation (\ref{heff-two-parts-generic})---which was
was the basis for our perturbative definition of a general surface near some
point---the coefficients in the two diagonal terms (such as $c$) control the curvature 
of the surface while the coefficients of the off-diagonal terms
(such as $a$ and $b$) control $\theta'/\theta$
(see equations (\ref{diff-theta-x}) and   (\ref{diff-theta-y})).
Further, as we have discussed, large `curvature coefficients' lead
to large $|\beta\rangle$ while large `theta variability coefficients' 
lead to large $|\Delta\alpha\rangle$.  The larger these coefficients
are, the larger $N$ must be to compensate, or higher order
terms would spoil the correspondence with the classical limit
we have built up.  Generally speaking, the factorization of eigenstate
property in equation (\ref{parallel}) fails when curvatures are too large at a given $N$
(since $|\beta\rangle$ becomes large).  On the other hand, when
the noncommutativity varies too quickly, the Poisson brackets
involving it (such as $\{N\theta,f\}$) will turn out to be too large.
 

\section{Area and minimal area surfaces}
\label{sec:area}

\begin{figure}
\includegraphics[width=6.5in]{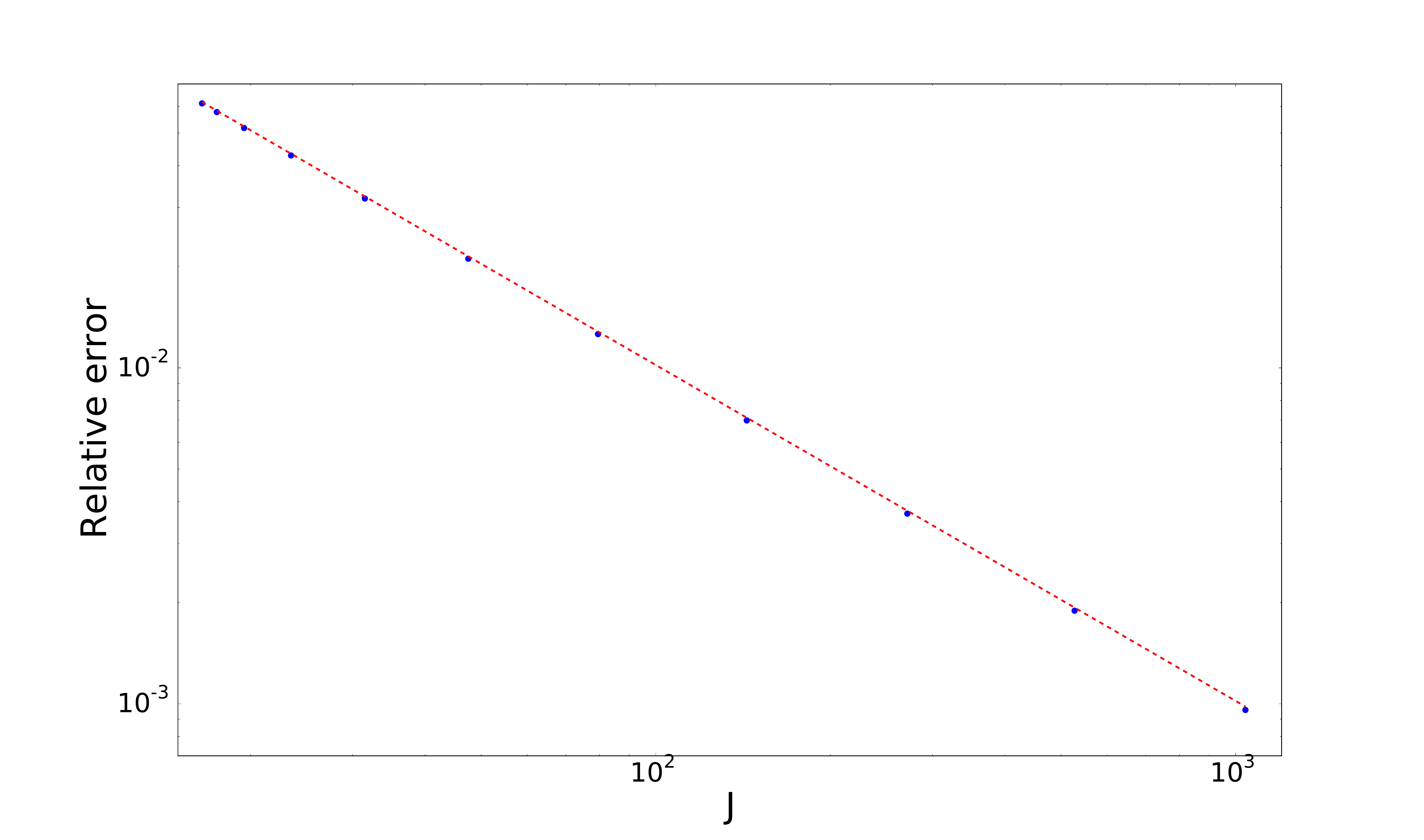}
\caption{Relative error in the noncommutative area as given in equation (\ref{area}) 
compared to the classical area, for an ellipsoid with major axes 6, 3 and 1. 
The error falls off with $J$ like $J^{-1}$; a best fit line, $1.02/J$, is shown to guide the
eye.}
\label{f5}
\end{figure}

In equation (\ref{theta-general}), we introduced an operator whose
expectation value in a coherent state is the local noncommutativity $\theta$.  
The noncommutativity $\theta$ has units of length-squared, and it can be interpreted as
the area of a single noncommutative `cell'.  This is similar to thinking of phase space
as made up of elementary cells whose area is $\hbar$.  In string theory, where a noncommutative
surface is made up of lower dimensional D-branes `dissolved' in the surface, we can
think of $\theta$ as the area occupied by a single D-brane, or, equivalently, the inverse
of the D-brane density.  If we divide the surface into $N$ noncommutative cells,
adding up the areas of all these cells we should get
the total area of the surface.  This is in fact borne out here, as the operator $\Theta$ introduced
in equation (\ref{theta-general}) has a second role: its trace seems to correspond to the area of
the surface\footnote{Factor of $2\pi$ can be arrived at by considering the round sphere.
Since our matrices $X_i$ are the SU(2) generators scaled by J, the more usual factor of $4\pi/N$
is multiplied by $J \approx N/2$.}
\be
A = 2\pi~ \mathrm{Tr} ~\Theta = 2\pi~ \mathrm{Tr} ~\sqrt {-\sum_{i,j} [X_i, X_j]^2}~.
\label{area}
\ee
Numerical evidence that this formula holds in is shown in figure \ref{f5}.

Consider now minimal area surfaces.  If we parametrize our emergent surface with
coordinates $\sigma_a$ and define the pullback metric on this surface:
\be
g_{ab} = \sum_{i=1}^3 ~\frac{\partial x_i}{\partial \sigma_a}~\frac{\partial x_i}{\partial \sigma_b}~,
\ee
(locally) minimal surfaces are solutions to the equations
\be
\Delta x_k(\sigma_a) = 0~,~~~k=1\ldots 3~,
\ee
where the Laplacian is, as usual
\be
\Delta = \frac{1}{\sqrt{g}} ~\frac{\partial}{\partial \sigma_a} ~\sqrt{g} g^{ab} ~\frac{\partial}{\partial \sigma_b} ~,
\ee
and where $g$ is the determinant of the metric $g_{ab}$.  

It is easy to check that these minimal surface equations can be written in terms of
the Poisson bracket (\ref{bracket}) as\footnote{This approach was used to
study matrix models for minimal area surfaces in \cite{Arnlind:2012cx}.}
\be
\sum_{i=1}^3\{x_i,\{x_i,x_k\}\} - \frac{1}{2}\sum_{i=1}^{3}~ \frac{\rho^2}{g}
\left\{x_i,\frac{g}{\rho^2}\right\}\{x_i,x_k\} = 0~.
\label{hoppe}
\ee
Let's now rewrite this equation in terms of $\theta$ (using equation (\ref{rho})):
\bear
\sum_{i=1}^3\{x_i,\{x_i,x_k\}\} - \frac{1}{2}\sum_{i=1}^{3}~ \theta^{-2}
\left\{x_i,\theta^2\right\}\{x_i,x_k\} &=& \\ \nn
 \sum_{i=1}^3\{x_i,\{x_i,x_k\}\} + \sum_{i=1}^{3}~ \theta
\left\{x_i,\theta^{-1}\right\}\{x_i,x_k\} &=& 0~,
\eear
or, in a more suggestive form (removing an overall factor of $\theta$),
\be
\sum_{i=1}^3 \{x_i,\theta^{-1}\{x_i,x_k\}\}= 0~.
\label{min-poisson}
\ee
This should be compared with the variation of our expression  for the area of
the noncommutative surface (\ref{area}):
\be
\frac{\partial A}{\partial X_1}  ~=~ \frac{1}{2} \left (~ \left [ X_2,  {\Theta}^{-1}[X_2,X_1]
+ [X_2,X_1] {\Theta}^{-1}  \right ]~+~(2\rightarrow 3) ~\right)~ = ~0~.
\label{min-commutators}
\ee
Taking an expectation value of equation (\ref{min-commutators}) w.r.t.
a coherent state, we obtain equation (\ref{min-poisson}), confirming that
the area of the noncommutative surface is indeed given by equation (\ref{area}).

Notice that this equation differs from that for a static configuration in
a generic matrix model (such as BFSS or IKKT), which is
\be
[X_i,[X_i,X_k]] = 0~.
\label{static}
\ee
This is because the Lagrangian for these matrix models contain a term of the form
$[X_i,X_j]^2$ which is the square of our operator $\Theta$.  When
considering minimum area surfaces in matrix models, when the noncommutativity
varies over the surface, the appropriate equation is not
(\ref{static}), but (\ref{min-commutators}), or more generally
\be
\Theta^{-1} [X_i,[X_i,X_k]] + [X_i,[X_i,X_k]] \Theta^{-1} +
[X_i,\Theta^{-1}][X_i,X_k] + [X_i,X_k][X_i,\Theta^{-1}]  = 0~,
\label{min-general}
\ee
which, in the large $N$ limit where ordering issues can be ignored, can be simplified to
\be
[X_i,[X_i,X_k]] ~+~ \Theta [X_i,\Theta^{-1}][X_i,X_k]  ~=~ 0
\label{min-general-2}
\ee
or
\be
[X_i,[X_i,X_k]] ~-~ \frac{1}{2} \Theta^{-2} [X_i,\Theta^2][X_i,X_k]  ~=~ 0~.
\label{min-general-3}
\ee
This last equation matches the original equation (\ref{hoppe}).
It is important to notice that the second term in the above equation (\ref{min-general-3})
has the same N-scaling as the first term: both are proportional to $N^{-2}$.  Thus,
this term cannot be neglected even in the large $N$ limit.

To gain more insight into the formula for the area of the surface,
we can examine the formula for the area in terms of the Poisson bracket:
\be
A = \int d^2\sigma \frac{\sqrt{g}}{N\theta}~\sqrt{\sum_{i,j}\{x^i,x^j\}}
~\rightarrow~ \int d^2\sigma \frac{\sqrt{g}}{\theta}~\sqrt{-[X_i,X_j]^2}~.
\label{nambu-goto}
\ee
The formula in equation (\ref{nambu-goto}) is essentially the bosonic part of the 
Nambu-Goto action for a string worldsheet.
This action is classically equivalent to the 
Schild action \cite{Schild:1976vq},
whose quantization via matrix regularization gives the IKKT model \cite{Ishibashi:1996xs}.
Equivalence of these two actions is proven by the standard method involving
an auxiliary field the inclusion of which removes the square root from the action 
\cite{Fayyazuddin:1997yf} (for a review, see \cite{Zarembo:1998uk}).
In the case of the correspondence between the Nambu-Goto and the Polyakov action,
this auxiliary field is the worldsheet metric.  Here, its role seems to be
linked to the local noncommutativity $\theta$.  This is not surprising: if the matrix
model is to be viewed as a quantization of the surface, we should be free to
pick any local noncommutativity we chose, so it can play the role of an
auxiliary field.  This
point of view provides a physical interpretation to the quantum equivalence of the 
IKKT and the nonabelian Born-Infeld model.

Finally, our computation allows us to write down the noncommutative Laplacian on
our emergent surface; it is, ignoring higher $1/N$-corrections 
\be
\Delta ~= ~\Theta^{-2} [X_i,[X_i,~\cdot~]] ~-~ \frac{1}{2}\Theta^{-4} [X_i,\Theta^2][X_i,~\cdot~] ~.
\ee
This equation could be the starting point for a study of the effects of varying
noncommutativity on noncommutative field theory.

\section{The torus}
\label{sec:torus}

Our construction has a natural extension to a toroidal surface embedded in 
flat three space.  Just as surfaces topologically equivalent to a sphere were
build by considering maps from the noncommutative sphere algebra, to make a torus
we use maps from the appropriate algebra.

Consider a  surface given by
\ba
x_1&= (R+r\cos u)\cos v~, \label{torus1}\\
x_2&= (R+r\cos u) \sin v~, \label{torus2}\\
x_3&= r \sin u~, \label{torus3}
\end{align}
where $u,v\in[0,2\pi]$ and $r<R$.  Now, consider the standard clock-and-shift matrices $U$
and $V$ that are usually used to define the noncommutative two-torus:
\ba
UV &= e^{2\pi i/N} VU~,\\
U_{kl}&=\delta_{kl}e^{2\pi i(k/N)}~, \\
V_{kl}&=\delta_{k_{\mathrm{mod}N},(l+1)_{\mathrm{mod}N}}~.
\end{align}
In the noncommutative torus, operators of the form $U^n V^m$ are 
associated with functions on the torus of the form $e^{i n u} e^{i n v}$.
To define the noncommutative torus embedded in $\mathbb{R}^3$ we thus simply
substitute $e^{iu} \rightarrow U$ and $e^{iv} \rightarrow V$ in
equations (\ref{torus1})-(\ref{torus3}), symmetrizing when necessary
to obtain hermitian matrices.  Numerical analysis shows that the 
resulting toroidal surface is smooth
and has the appropriate large $N$ limit (with $A_p$ decreasing for
large $N$ as $N^{-1/2}$, the surface approaching the classical shape
and the area of the surface well approximated by equation (\ref{area})).


Once we have obtained this particular toroidal surface, any other
surface with this topology (including surfaces with the same shape but
different local noncommutativity, for example uniform one) can be obtained by 
smooth maps in a way that parallels our discussion of spherical surfaces.
It would be interesting to consider a deformation which connects
the torus and the sphere and to examine what happens at the point
of topological transition in detail.

\section{Open questions and future work}
\label{future}

There are many questions which our work does not address.  

For example, one can ask if equation (\ref{area}) can be proven analytically, starting
with the definition of the surface from $\Heff$.  A reasonable start
for such a proof might be equation (\ref{nambu-goto}).
If we assume that
\be
\frac{1}{N}~\mathrm{Tr} ~~\cdot ~=~ \frac{1}{2\pi} \int d^2\sigma \frac{\sqrt{g}}{N\theta}~ \langle \alpha(\sigma) |~~ \cdot~~
| \alpha(\sigma) \rangle ~,
\label{trace}
\ee
we recover equation (\ref{area}).
Equation (\ref{trace}) is equivalent to 
\be
 \frac{1}{2\pi} \int d^2\sigma \frac{\sqrt{g}}{\theta}~|\alpha(\sigma)\rangle\langle\alpha(\sigma)|~=~
\boldsymbol{1}_N~.
\label{completness}
\ee
Above equation 
implies a relationship between the trace and the integral of the noncommutative surface
\be
\frac{1}{N}~\mathrm{Tr} ~~\leftrightarrow~~  \frac{1}{2\pi}\int d^2\sigma \frac{\sqrt{g}}{N\theta}~.
\ee

A completeness relationship such as (\ref{completness}) is 
necessary for the symbol map from operators to functions on the emergent
surface to have a unique inverse, which in turn is necessary
for the definition of the star product to make sense.  In principle, it
should be possible to prove such a completeness relationship starting
with equation (\ref{BD}).

In subsection \ref{subsection:coherent}, we briefly addressed the question of
the U(1) connection on the emergent D2-brane.  Extending this
approach should allow us to prove the equivalence of the nonabelian effective
action for D0-branes and the abelian effective action for a D2-brane.  More simply, it
should be possible to show the equivalence of the BPS conditions in these two scenarios.

It would be interesting to see how our set up could be extended to surfaces
which are not topologically equivalent to a sphere or a torus.  It should be possible, for example,
to find matrix triplets $X_i$ which correspond to emergent surfaces with a
larger number of handles---and
for which the large $N$ limit we describe holds.  One could check, for example,
whether the noncommutative surfaces given in \cite{2007arXiv0711.2588A} have a 
large $N$ limit in the sense in which we define it here.  Further, it would be
interesting to see how our toroidal construction in section \ref{sec:torus}
is related to that in  \cite{2007arXiv0711.2588A}.


Finally, there are many generalizations of equation (\ref{BD}) that would
be interesting to explore, including generalizations to higher dimensions
(both of the embedding space and the emergent surface) and those to curved
embedding space.  One could also consider Lorentzian signature models,
which would be useful in the context of recent progress in cosmology
arising from matrix models, as in \cite{Kim:2011cr}.


\section*{Acknowledgments}

This research was partially
funded by the Natural Sciences and Engineering Research Council of Canada (NSERC).
PSG's research was partially funded by Fonds de recherche du Qu\'{e}bec---Nature et technologies, 
and a Walter C. Sumner Memorial Fellowship.

\bibliographystyle{JHEP}
\bibliography{new}

\end{document}